\documentclass[twocolumn,aps,superscriptaddress,nofootinbib]{revtex4-1}
\usepackage{pstricks}
\usepackage{graphicx}
\usepackage{caption}
\usepackage{subcaption}
\usepackage{amsmath}
\usepackage{amssymb}
\usepackage{slashed}
\usepackage{xcolor}
\usepackage[normalem]{ulem}

\usepackage[colorlinks]{hyperref}  

\graphicspath{{./plots/}}

\begin{document}
\title{Cabibbo suppressed single pion production off the nucleon induced by 
antineutrinos}
\author{M. Benitez Galan}
\email{mbenitezgalan@ugr.es}
\affiliation{Departamento de F\'isica
At\'omica, Molecular y Nuclear, 
Facultad de Ciencias, 
Universidad de Granada, E-18071, Granada,
  Spain}
\author{M. Rafi Alam}
\email{rafi.alam.amu@gmail.com}
\affiliation{Department of Physics,
Aligarh Muslim University, Aligarh-202 002,
India}
\author{I. Ruiz Simo}
\email{ruizsig@ugr.es}
\affiliation{Departamento de F\'isica
At\'omica, Molecular y Nuclear and Instituto
Interuniversitario Carlos I de F\'isica
Te\'orica y Computacional, 
Facultad de Ciencias, 
Universidad de Granada, E-18071, Granada,
  Spain}

\begin{abstract}
In this work we study the $\Sigma\pi$ and $\Lambda\pi$ production off
free nucleons driven by the strangeness-changing weak charged
current. We calculate the total cross sections for all possible
channels and estimate the flux-averaged total cross sections for
experiments like MiniBooNE, SciBooNE, T2K, and Minerva.  The model is
based on the lowest order effective SU(3) chiral Lagrangians in the
presence of an external weak charged current and contains Born and the
lowest-lying decuplet resonant mechanisms that can contribute to these
reaction channels.  We also compare and discuss our results with
others following similar and very different approaches.
\end{abstract}

\maketitle

\section{Introduction}\label{sect:intro}
The neutrino and antineutrino-nucleus cross sections are necessary
inputs for the analyses of the neutrino scattering and oscillation
experiments \cite{Alvarez-Ruso:2017oui,
  Alvarez-Ruso:2014bla,Katori:2016yel,
  Mosel:2016cwa,Mosel:2019vhx}. One of the main ingredients in the
(anti)neutrino-nucleus cross sections is the primary
(anti)neutrino-nucleon interaction model. It is very important that
these models provide accurate predictions when compared with
experimental data on nucleon targets, before embedding these
elementary interactions within the nuclear medium, where relevant
nuclear effects may distort the final signal in experiments.  In the
few GeV energy regions, where most of the
present~\cite{Drakoulakos:2004gn, Abe:2011ks,Chen:2007ae} and
future~\cite{Abe:2015zbg, Acciarri:2016crz,Acciarri:2015uup}
oscillation experiments take data, single pion production channels may
play a crucial role.

The Cabibbo enhanced single pion production off nucleons is a
long-standing theoretical process that has been studied
\cite{Dennery:1962zz, Rein:1980wg,
  Wilkinson:2014yfa,Hernandez:2013jka,Paschos:2000be,
  Lalakulich:2010ss,Kuzmin:2006dh,Wu:2014rga,
  Leitner:2008wx,Kabirnezhad:2017jmf,Mosel:2017nzk,
  Ivanov:2012fm,Nikolakopoulos:2018gtf,Barbero:2013eqa,
  Martini:2009uj,Hernandez:2007qq,
  Sobczyk:2018ghy,Yao:2018pzc,Yao:2019avf,
  AlvarezRuso:1998hi,Leitner:2008ue,Ahmad:2006cy,
  Alam:2015gaa,Gonzalez-Jimenez:2016qqq,Lalakulich:2005cs,
  Lalakulich:2006sw,Gershtein:1980vd} and measured
\cite{Budagov:1969pw,Radecky:1981fn,Kitagaki:1986ct,
  Krenz:1977sw,Lee:1976wr,Grabosch:1988gw,McGivern:2016bwh,
  Kurimoto:2009wq,Barish:1975bw,Erriquez:1978yc,Abe:2016aoo,
  Bolognese:1979gf,Derrick:1980xw,Baker:1980pj,Hasert:1975sv,
  Pohl:1979fw,Allasia:1983qh,Coplowe:2020yea,
  AguilarArevalo:2010bm,Eberly:2014mra,
  Aliaga:2015wva,AguilarArevalo:2010xt,Rodriguez:2008aa} since many
decades ago up to date. However, its Cabibbo suppressed counterpart,
where a pion is produced along with a $S=-1$ hyperon ($\Sigma$ or
$\Lambda$) in the final state, is a scarcely studied set of reactions.

In the previous
works~\cite{Ren:2015bsa,Wu:2013kla,Finjord:1975zy,Dewan:1981ab},
different approaches have been followed. In Ref. \cite{Ren:2015bsa} a
coupled-channel chiral unitary approach is used to dynamically
generate the $\Lambda(1405)$ resonance, which plays a major role in
the $\pi\Sigma$ reaction channel. In Refs.
\cite{Wu:2013kla,Finjord:1975zy} a non-relativistic 3-quark model,
effective $V-A$ theory with experimental form factors, and the
relativistic quark model with harmonic interaction of Feynman,
Kislinger and Ravndal \cite{Feynman:1971wr} are used to calculate the
cross section for $\Sigma^{*0}(1385)$ resonance production off proton,
among other channels. Finally, in Ref.  \cite{Dewan:1981ab} a model
with background or Born terms is used to calculate a plethora of
reactions producing strange particles, in particular the $\pi Y$
production channel, but explicitly excluding $N^*$ and $Y^*$ exchange
mechanisms.

The kind of reactions studied in this work can only be induced by
antineutrinos, due to the selection rule for the strangeness-changing
weak charged current, $\Delta S=\Delta Q=-1$, for the hadrons. Given
that the strangeness-changing weak current changes an $u$ quark into a
$s$ quark (or a $\bar{s}$ antiquark into an $\bar{u}$ one), there are
also the selection rules $\Delta I=\frac12$ and $\Delta
I_{z}=-\frac12= \frac{\Delta Q}{2}$, where $(I,I_z)$ are the strong
isospin and its third component.

Though the present work centered around strangeness changing pion
production, the hyperon produced in the final state holds an added
advantage.  For instance, the inclusive hyperon ($\Lambda$ or
$\Sigma$) production below the energy threshold for associated $KY$
production is going to be dominated by the quasielastic (QE) hyperon
production channel \cite{Singh:2006xp,Kuzmin:2008zz,
  Wu:2013kla,Alam:2014bya,Fatima:2018wsy, Sobczyk:2019uej} and by the
reactions studied in this work. In particular, the direct $\Sigma^{+}$
production in QE hyperon reactions off nucleons is not allowed; the
final appearance of $\Sigma^{+}$ particles in reactions taking place
off nuclear targets is due to the final-state interactions (FSI) or
re-scattering experienced by the other hyperons inside the nucleus
\cite{Singh:2006xp,Fatima:2018wsy,Alam:2014bya}.  However, in the
inelastic$(\Delta S =-1)$ channel, $\Sigma^{+}$ can be produced in
primary antineutrino interaction off protons(for a complete list of
final states, please see Sec.~\ref{sec:formalism}), which is expected
to be dominant source of $\Sigma^{+}$ production below the $KY$
threshold.  Also, a direct consequence of FSI and nuclear effects is
the absorption of produced (primary) pions on a large scale; however,
the secondary pions produced from hyperon decay will not suffer a
strong absorption thanks to the long lifetime of hyperons.
 
In this work, we developed a model for (anti)neutrino-induced $\pi Y$
production on the nucleon induced by the charged current
interactions. The present model is largely based on the models that
have been well tested in the past, like in
$K$-production~\cite{RafiAlam:2010kf,Alam:2012zz,Alam:2012ry},
$\pi$-production~\cite{Hernandez:2007qq} etc. While the non-resonant
mechanism relies on the chiral Lagrangian and SU(3) flavor symmetry,
the resonant mechanism involves both non-strange ($\Delta(1232))$ and
strange ($\Sigma^*(1385))$ resonances.

The structure of this work is as follows: in Sect \ref{sec:formalism}
we discuss the formalism in detail; in Sect \ref{sec:results} we
present our results; and finally, in Sect \ref{sec:conclusions} we
summarize our findings.

\section{Formalism}\label{sec:formalism}
In this work we are interested in the following set of antineutrino
induced reactions
\begin{eqnarray}
 \bar{\nu}_l(k) + N(p) &\longrightarrow&
 l^{+}(k^\prime) + \pi(p_m) + Y(p_Y),
 \nonumber\\
 \label{eq:reaction}
\end{eqnarray}
where $N$ can be either a proton or neutron, $Y$ is a $\Sigma$ or
$\Lambda$ hyperon, and the four-momenta of particles are given in
parentheses.  For induced reactions off protons, the allowed $Y\pi$
final states are $\Lambda\pi^0$, $\Sigma^0\pi^0$, $\Sigma^+\pi^-$ and
$\Sigma^-\pi^+$; while for the neutron channel the possibilities are
$\Lambda\pi^-$, $\Sigma^0\pi^-$ and $\Sigma^-\pi^0$.

\begin{figure*}
\begin{subfigure}[h]{0.55\textwidth}
\includegraphics[width=0.95\textwidth,height=.65\textwidth]{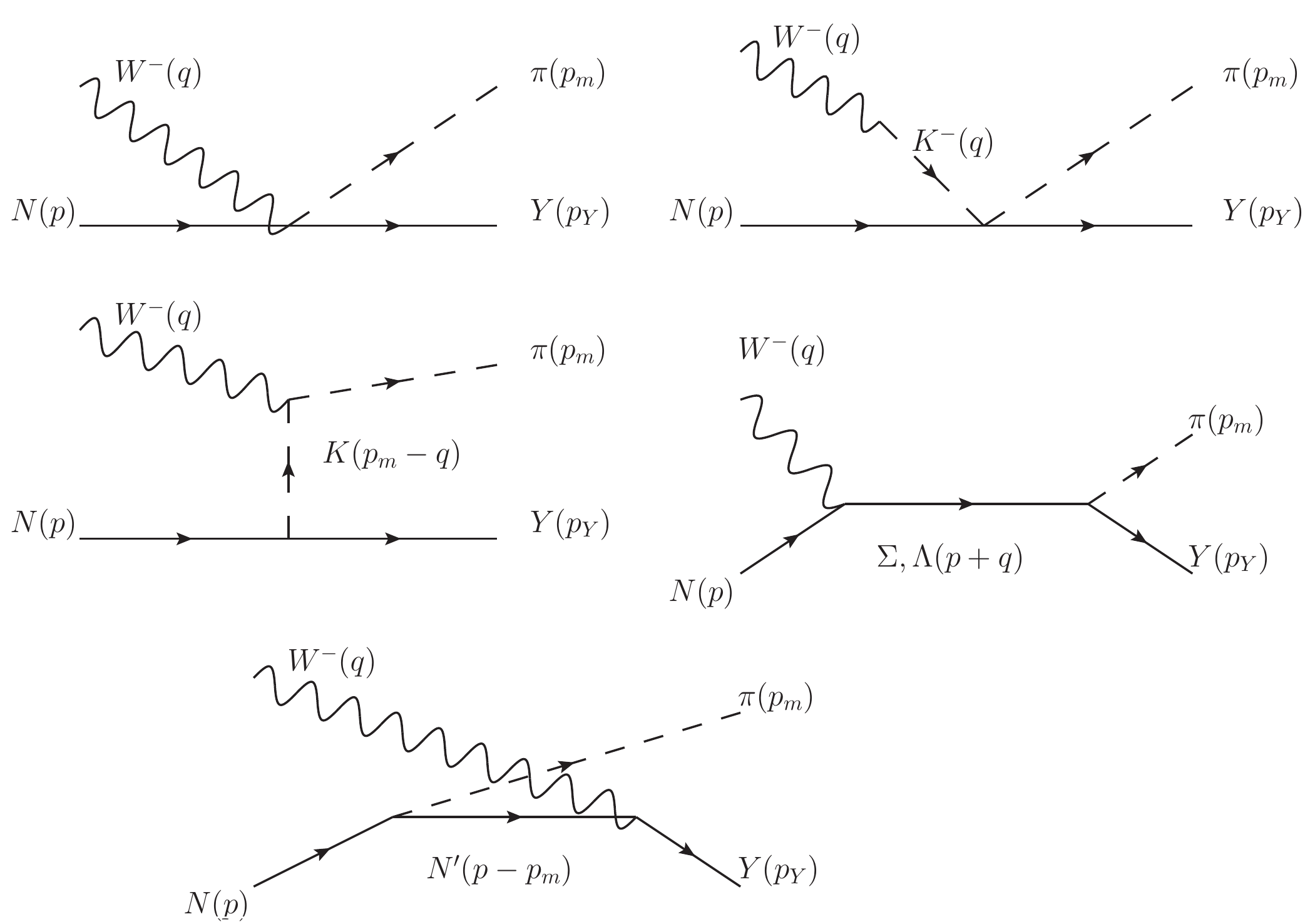}
\caption{Background or Born terms of our model. From top to bottom and
  from left to right, we find the contact term (CT), the kaon pole
  (KP), the kaon-in-flight (KF), the s-channel $\Sigma$ and $\Lambda$
  (s-$\Sigma$ and s-$\Lambda$) and the u-channel $N$ (u-$N$) diagrams,
  respectively.}\label{fig:background}
\end{subfigure}
\begin{subfigure}[h]{0.44\textwidth}
\includegraphics[width=0.95\textwidth,height=.65\textwidth]{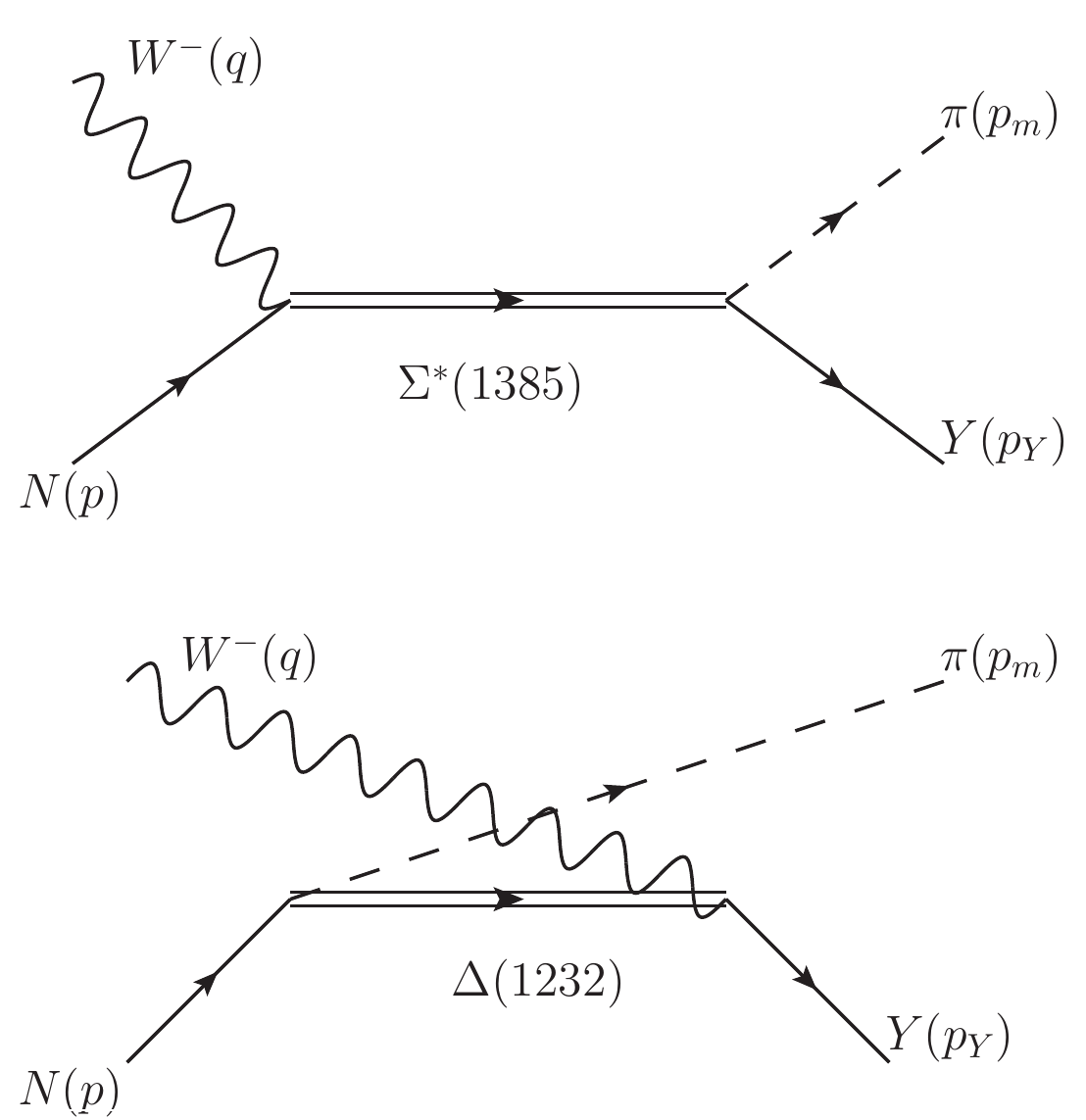}
\caption{Resonance diagrams included in our model.  The s-channel
  $\Sigma^*(1385)$ diagram is shown in the upper figure, while the
  u-channel $\Delta(1232)$ diagram is depicted in the lower figure. }
\label{fig:resonances}
\end{subfigure}
\caption{Feynman diagrams included in our model for the Cabibbo
  suppressed $\pi Y$ production process off nucleons induced by
  antineutrinos.}
\label{fig: model}
\end{figure*}

Our model, shown in Fig. \ref{fig: model}, is very similar to that of
Ref. \cite{Dewan:1981ab}, but also includes the lowest lying decuplet
resonances like $\Delta(1232)$ and $\Sigma^*(1385)$ as explicit
degrees of freedom (shown in Fig. \ref{fig:resonances}), in the line
of previous works such as those of
Refs. \cite{RafiAlam:2010kf,Alam:2012zz,
  Alam:2012ry,RafiAlam:2019rft}. We use effective $V-A$
strangeness-changing weak charged current with vector and axial-vector
form factors for the $N-Y^\prime$ transitions.  The vector form
factors are related to the electromagnetic nucleon form factors using
the Cabibbo theory, i.e, assuming that the strangeness-changing weak
vector current belongs to an SU(3) octet of flavor currents.  For the
axial-vector currents, $D$-type (symmetric) and $F$-type
(antisymmetric) couplings arise between two octets $\left\lbrace 8
\right\rbrace\otimes\left\lbrace 8 \right\rbrace$ that are connected
through a SU(3) octet axial current.  Whereas, the $q^2$-dependence is
introduced by assuming a similar form for both $D$ and $F$ couplings,
taken to be of dipole form~\cite{RafiAlam:2019rft,Singh:2006xp}.  For
the $\pi NN^\prime$ and $\pi Y Y^\prime$ strong vertices we assume
pseudo-vector couplings with the derivative of the pseudo-scalar meson
field. These assumptions are fully consistent with the lowest order
baryon-meson chiral Lagrangians in the presence of a weak charged
external current, as discussed in \cite{Scherer:2012xha}.

\subsection{Total cross section}
\label{subsec:totalxsect}

The unpolarized differential cross section corresponding to
eq. (\ref{eq:reaction}) is
\begin{eqnarray}
 &&d^9\sigma= \delta^4(p+q-p_Y-p_m)
 \frac{1}{(2\pi)^5\,4ME_{\bar{\nu}}}\nonumber\\
 &&\frac{d^3k^\prime}{2
 E^\prime_l(\mathbf{k}^\prime)}\;
 \frac{d^3p_m}{2E_m(\mathbf{p}_m)}\;
 \frac{d^3p_Y}{2E_Y(\mathbf{p}_Y)} 
 \overline{\sum}\sum
 \left|\mathcal{M}\right|^2,\nonumber\\
 \label{eq: d9xsect}
\end{eqnarray}
where the matrix element $\mathcal{M}$ is
 \begin{equation}\label{eq:red_matelem}
-i\mathcal{M}=-i\,\frac{G_F}{\sqrt{2}}\; \ell_\mu J^\mu_H,
\end{equation}
with $G_F=\frac{\sqrt{2}g^2}{8M^2_W}=1.1664\times 10^{-5}$ GeV$^{-2}$
as the Fermi coupling constant and $\ell_\mu(J^\mu_H)$ is the lepton
(hadron) current.  For the final calculations, we use $J^\mu_H$ given
as the sum of the hadronic currents of
eqs. (\ref{eq:currentCT}-\ref{eq:current_uN}) and
(\ref{eq:current_ssigmastar}-\ref{eq:current_delta}). The symbol
$\overline{\sum}\sum \left|\mathcal{M} \right|^2$ stands for the sum
over final fermion spins and average over initial ones if these are,
on average, unpolarized.  In the present calculations, we take initial
nucleons as unpolarized; however, antineutrinos are fully polarized,
which leads to
\begin{equation}\label{eq:squared_mat_elem}
 \overline{\sum}\sum \left|\mathcal{M} 
\right|^2=2\,G^2_F\, L^{\mu\nu}(k,k^\prime)
\sum_{\lambda_N,\lambda_Y} J^H_\mu 
(J^H_\nu)^{*}.
\end{equation} 
In the above expression, $L^{\mu\nu}(k,k^\prime)$ is the lepton tensor
\begin{eqnarray}
L^{\mu\nu}(k,k^\prime)&=&
 k^\mu k^{\prime\nu}+k^\nu k^{\prime\mu}-
 g^{\mu\nu}(k\cdot k^\prime)\nonumber\\ 
&& - i\, \epsilon^{\mu\nu\alpha\beta}k_{\alpha}
 k^\prime_{\beta},\label{eq:lepton_tensor}
\end{eqnarray}
with $\epsilon^{0123}=1$. Finally, the sum over the spins of the
initial and final baryons $(\lambda_{N,Y})$ gives rise to traces over
chains of Dirac matrices, of the form
\begin{eqnarray}
W_{\mu\nu}&=& 
\sum_{\lambda_N,\lambda_Y} J^H_\mu 
(J^H_\nu)^{*}=\sum_{\lambda_N,\lambda_Y}
\left[\bar{u}_{\lambda_Y}(\mathbf{p}_Y)j_{\mu}
u_{\lambda_N}(\mathbf{p})\right]\nonumber\\
&&\left[ \bar{u}_{\lambda_N}(\mathbf{p})
\gamma^0 j^\dagger_{\nu} \gamma^0
u_{\lambda_Y}(\mathbf{p}_Y)\right]=\nonumber\\
&=&{\rm Tr}\left[ j_\mu (\slashed{p}+M) 
\gamma^0 j^\dagger_{\nu} \gamma^0
(\slashed{p}_Y+M_Y)\right],
\label{eq:hadron_tensor}
\end{eqnarray}
where $j_\mu$ is the total hadron current $J^H_\mu$, but without Dirac
spinors as given in eqs.  (\ref{eq:currentCT}-\ref{eq:current_uN}) and
(\ref{eq:current_ssigmastar}-\ref{eq:current_delta}).  For the
calculation of Dirac traces, we have used the Mathematica package
Feyncalc~\cite{Shtabovenko:2020gxv,
  Shtabovenko:2016sxi,Mertig:1990an}.

The eq. (\ref{eq: d9xsect}) can be further solved with the help of the
$\delta$-function. The delta integration then fixes the cosine of the
polar angle theta $(\theta_m^0 = \cos^{-1}[\hat q \cdot \hat p_m])$:
\begin{equation}\label{eq:costhetasol}
\cos\theta^0_m=\frac{M^2_Y + \mathbf{q}^2+\mathbf{p}^2_m-
(M+q^0-E_m)^2}{2\left|\mathbf{q} \right| \left|\mathbf{p}_m \right|},
\end{equation}
and the eq. (\ref{eq: d9xsect}) thus reduces to, 
\begin{eqnarray}
&&d^5\sigma=\frac{1}{(2\pi)^5\,4ME_{\bar{\nu}}}\; 
\frac{\left| \mathbf{k}^\prime \right|}{8\left| \mathbf{q} \right|}\;
\overline{\sum}\sum\left| \mathcal{M} \right|^2 \nonumber\\
&&\Theta(1-\cos^2\theta^0_m)\,dE^\prime_l\,d\Omega_{\hat{k}^\prime}\,
dE_m\,d\phi_m,\label{eq:d5xsect}
\end{eqnarray}
where $\phi_m$ is the azimuthal angle of the three-momentum of the
$\pi$ meson on the reaction plane measured with respect to the
$\bar{\nu}-l^{+}$ scattering plane. The step function ($\Theta$) puts
a constraint on the cosine of theta ($\theta_m^0$).

Finally, integrating eq. (\ref{eq:d5xsect}) with respect to all the
variables for a fixed antineutrino energy $E_{\bar{\nu}}$, we obtain
\begin{eqnarray}
&&\sigma(E_{\bar{\nu}})=\frac{1}{(2\pi)^5\,4ME_{\bar{\nu}}}\;
\int d\Omega_{\hat{k}^\prime}\,\int^{E^\prime_{l{\rm max}}}_{m_l}
dE^\prime_l\; \frac{\left| \mathbf{k}^\prime \right|}{8\left| \mathbf{q} \right|}
\nonumber\\
&&\int^{E^{\rm max}_{m}}_{m_\pi} dE_m\;\Theta(1-\cos^2\theta^0_m)
\int^{2\pi}_{0}d\phi_m\, \overline{\sum}\sum\left| \mathcal{M} \right|^2.
\nonumber\\
\end{eqnarray}
For the upper limits of integration in the energies of the final
lepton and the $\pi$ meson, we have chosen $E^\prime_{l{\rm max}}=
E_{\bar{\nu}}+M-M_Y-m_\pi$ and $E^{\rm max}_m=
E_{\bar{\nu}}-E^\prime_l+M-M_Y$.

\subsection{Born terms model}\label{subsec:born}
Following Refs. \cite{Scherer:2012xha,Gasser:1984gg} we can write the
lowest order chiral Lagrangian in the SU(3) flavor scheme for mesons
in the presence of an external weak charged current as
\begin{equation}\label{eq:lag_mesons}
\mathcal{L}^{(2)}_{M}= \frac{f^2_\pi}{4} {\rm Tr}\left[ 
D_{\mu}U(D^{\mu}U)^\dagger \right] + \frac{f^2_\pi}{4} {\rm Tr}\left[
\chi U^\dagger + U\chi^\dagger \right],
\end{equation}
where $f_\pi=93$ MeV is the pion decay constant, $U$ is the SU(3)
representation of the pseudo-scalar octet meson fields
\begin{eqnarray}
U(x)&=&\exp\left(i\frac{\phi(x)}{f_\pi} \right)\nonumber\\
\phi(x)&=&\left(
\begin{array}{ccc}\label{eq:meson_fields}
\pi^0+\frac{\eta}{\sqrt{3}} & \sqrt{2}\pi^{+} & \sqrt{2}K^{+} \\
 \sqrt{2}\pi^{-} & -\pi^0+\frac{\eta}{\sqrt{3}} & \sqrt{2}K^{0} \\
\sqrt{2}K^{-} & \sqrt{2}\bar{K}^{0} & -\frac{2}{\sqrt{3}}\eta
\end{array}
\right).
\end{eqnarray}
$D_{\mu}U$ is the covariant derivative, given by
\begin{equation}
D_{\mu}U=\partial_{\mu}U -i r_\mu U + i U l_\mu,
\end{equation}
where $l_{\mu}$ and $r_{\mu}$ are left and right-handed external
currents coupled to the meson fields. In the particular case of the
weak charged current, these currents are:
\begin{equation}
r_{\mu}=0 \qquad l_{\mu}=-\frac{g}{\sqrt{2}}\left( 
W^{+}_{\mu} T_{+} + W^{-}_{\mu} T_{-} \right),
\end{equation}
with $W^{\pm}_{\mu}$ the weak vector boson fields, $g$ the weak
coupling constant, and $T_{\pm}$ the $3\times3$ matrices containing
the Cabibbo-Kobayashi-Maskawa matrix elements relevant for the three
flavor scheme,
\begin{equation}
T_{+}=\left(
\begin{array}{ccc}
0 & V_{ud} & V_{us} \\
0 & 0 & 0 \\
0 & 0 & 0
\end{array}
\right); \quad
T_{-}=\left(
\begin{array}{ccc}
0 & 0 & 0 \\
V_{ud} & 0 & 0 \\
V_{us} & 0 & 0
\end{array}
\right).
\end{equation}
Finally, in eq. (\ref{eq:lag_mesons}), the symbol ${\rm Tr}$ denotes a
trace over flavor space. The second term in eq. (\ref{eq:lag_mesons})
is not relevant for our study. It incorporates the explicit breaking
of chiral symmetry due to the finite quark masses.  With the
Lagrangian given in eq. (\ref{eq:lag_mesons}) we can obtain the
relevant $WK\pi$ and $W\bar{K}$ vertices necessary for the KP and KF
diagrams shown in Fig.  \ref{fig:background}.

The lowest order interaction between the octet baryons, the octet
meson and the weak external current can also be introduced following
Ref. \cite{Scherer:2012xha} as
\begin{eqnarray}
 &&\mathcal{L}^{(1)}_{MB}=
 {\rm Tr}\left[\bar{B}(i\slashed{D}
 -M)B\right]\nonumber\\
 &+&
 \frac{D}{2}
 {\rm Tr}\left[\bar{B}\gamma^\mu
 \gamma_5\left\lbrace u_\mu,B
 \right\rbrace\right]
 +
 \frac{F}{2}
 {\rm Tr}\left[\bar{B}\gamma^\mu
 \gamma_5\left[ u_\mu,B
 \right]\right],\nonumber\\\label{eq:lag_mb}
\end{eqnarray}
where $B(x)$ is the SU(3) representation of the baryon fields
\begin{equation}\label{eq:baryon_matrix}
B=\left(
\begin{array}{ccc}
\frac{1}{\sqrt{2}}\Sigma^0 + 
\frac{1}{\sqrt{6}}\Lambda & \Sigma^+ & p \\
\Sigma^{-} & -\frac{1}{\sqrt{2}}\Sigma^0 
+ \frac{1}{\sqrt{6}}\Lambda & n \\
\Xi^{-} & \Xi^0 & -\frac{2}{\sqrt{6}}\Lambda
\end{array}
\right).
\end{equation}
The covariant derivative of the baryon fields is given in terms of the
connection $\Gamma_{\mu}$ as
\begin{equation}
 D_{\mu}B=\partial_\mu B + \left[ \Gamma_\mu,B
 \right],
\end{equation}
with
\begin{equation}\label{eq:connection}
 \Gamma_\mu=\frac12\left[ u^\dagger\left(
 \partial_\mu-ir_\mu\right)u + u\left(
 \partial_\mu-il_\mu\right)u^\dagger\right].
\end{equation}
In eq. (\ref{eq:connection}) we have introduced
$u=\sqrt{U}=\exp\left(i\frac{\phi(x)}{2f_\pi} \right)$. Also, in
eq. (\ref{eq:lag_mb}), the definition of the so-called vielbein,
$u_\mu$, is given by
\begin{equation}\label{eq:vielbein}
 u_\mu=i\left[u^\dagger\left( \partial_\mu 
 -ir_\mu\right)u - u\left( \partial_\mu 
 -il_\mu\right)u^\dagger\right].
\end{equation}
In eq. (\ref{eq:lag_mb}), $M$ represents the baryon mass matrix in the
exact SU(3) limit with $D(=0.804)$ and $F(=0.463)$ as the symmetric
and antisymmetric couplings, respectively.  The two independent
couplings appear because in the Clebsch-Gordan series expansion of two
SU(3) octets $\left\lbrace 8 \right\rbrace \otimes\left\lbrace 8
\right\rbrace$, the $\left\lbrace 8 \right\rbrace$ representation is
contained twice. These couplings can be measured from the baryon
semileptonic decays within the Cabibbo model \cite{Cabibbo:2003cu}.
The Lagrangian of eq. (\ref{eq:lag_mb}) allows to extract all the
necessary vertices $NYK$, $NYK\pi$, $NYW\pi$, and the leading order
vector and axial-vector terms for the $N-Y$ strangeness-changing weak
transitions for the diagrams depicted in Fig. \ref{fig:background}.
The latter can be written as
\begin{widetext}
\begin{eqnarray}
 \left\langle Y(p^\prime_Y)\right| V^\mu 
 \left| N(p) \right\rangle &=& 
 \bar{u}_Y(\mathbf{p}^\prime_Y)\left[
 f^{NY}_1(q^2) \gamma^\mu + i 
 \frac{f^{NY}_2(q^2)}{M+M_Y}
 \sigma^{\mu\nu}q_{\nu} + 
 \frac{f^{NY}_3(q^2)}{M+M_Y}q^\mu
 \right]u_N(\mathbf{p}) \label{eq:vector_matelem} \\
 \left\langle Y(p^\prime_Y)\right| A^\mu 
 \left| N(p) \right\rangle &=& 
 \bar{u}_Y(\mathbf{p}^\prime_Y)\left[
 g^{NY}_1(q^2) \gamma^\mu\gamma_5 + i 
 \frac{g^{NY}_2(q^2)}{M+M_Y}
 \sigma^{\mu\nu}\gamma_5q_{\nu} + 
 \frac{g^{NY}_3(q^2)}{M+M_Y}q^\mu\gamma_5
 \right]u_N(\mathbf{p}). \label{eq:axial_matelem}
 \end{eqnarray}
\end{widetext}
where $(g^{NY}_i)f^{NY}_i$, $i=1,2,3$ are the (axial-)vector form
factors.  The Lagrangian of eq. (\ref{eq:lag_mb}) provides the values
for the vector and axial couplings (form factors at $q^2=0$)
$f^{NY}_1(0)$ and $g^{NY}_1(0)$, but not for the others, which may
appear at higher orders of the chiral expansion. However, using
symmetry arguments, one can get rid of some of them.  For example, the
weak electricity ($g^{NY}_2(q^2)$) and the scalar ($f^{NY}_3(q^2)$)
form factors transform as second-class currents \cite{Weinberg:1958ut}
under G-parity and are neglected for present calculations~\footnote{We
assume that G-parity is a good quantum number for the strong
interactions and that in the Standard Model there are no second-class
currents. Therefore, from here onward we neglect the contribution of
$g_2$ and $f_3$.  For an exhaustive discussion and implications of
their effects in some observable if second-class currents are sizable,
the reader is referred to Ref. \cite{Fatima:2018tzs} and references
therein.}.  In the present scheme the most standard way to obtain the
$f_2(0)$ couplings is to include the relevant pieces of the next
higher order meson-baryon chiral Lagrangian \cite{Oller:2006yh} and to
match the low energy constants to well-known $f_2(0)$ transition form
factors, which can be obtained from Table I of
Ref. \cite{Cabibbo:2003cu}.

Similar results could have been achieved by invoking exact SU(3)
symmetry and the hypothesis that the weak vector currents and the
electromagnetic one belong to the same octet of current operators of
the SU(3) group. As the octet $\left\lbrace 8 \right\rbrace$
representation appears twice in the Clebsch-Gordan series for the
tensor product of two octets
\begin{equation}\label{eq:CG_series}
 \left\lbrace 8 \right\rbrace\otimes
 \left\lbrace 8 \right\rbrace = 
 \left\lbrace 1 \right\rbrace \oplus 
 \left\lbrace 8 \right\rbrace \oplus
 \left\lbrace 8^\prime \right\rbrace \oplus
 \left\lbrace 10 \right\rbrace \oplus
 \left\lbrace \overline{10} \right\rbrace \oplus
 \left\lbrace 27 \right\rbrace,
\end{equation}
this means that \emph{any octet operator connecting two octet baryons
has two independent irreducible matrix elements}.  Therefore, it is
necessary to explicitly calculate two independent matrix elements for
an octet operator. Later, using the SU(3) Wigner-Eckart theorem, all
the non-vanishing matrix elements between octet states connected
through an octet current operator can be related through the SU(3)
Clebsch-Gordan coefficients, which can be found in
Ref. \cite{McNamee:1964xq}, with the previous explicitly calculated
two matrix elements. In the case of the octet of vector currents,
these two irreducible matrix elements can be written in terms of the
proton and neutron electromagnetic current matrix elements,
$\left\langle p\right| J^\mu_{\rm em}\left|p\right\rangle$ and
$\left\langle n\right| J^\mu_{\rm em}\left|n\right\rangle$.  This
facilitates us to express all the $N \rightleftharpoons Y$ transition
vector form factors in terms of those, $f^{p,n}_{1,2}(q^2)$, of the
electromagnetic interaction, that is well measured.  They are
summarized in Table \ref{tab:vector_ff}, and for present work we use
the Galster parameterization \cite{Galster:1971kv} for the
electromagnetic form factors.

\begin{table*}[ht]
\begin{tabular}{|c|c|c|c|}
\hline
 $i=1,2$ &  $Y=\Lambda$ & $Y=\Sigma^0$ & $Y=\Sigma^-$\\
 \hline
 $f^{pY}_{i}(q^2)$ &$-\sqrt{\frac32}f^p_i(q^2)$&
 $-\frac{1}{\sqrt{2}}\left( f^p_i(q^2)+
 2f^n_i(q^2)\right)$& $0$ \\
 \hline
 $f^{nY}_{i}(q^2)$ & $0$ & $0$ & $-\left(
 f^p_i(q^2)+2f^n_i(q^2)\right)$\\
 \hline
\end{tabular}
\caption{Dirac and Pauli vector form factors for the weak
  strangeness-changing transitions considered in this
  work.}\label{tab:vector_ff}
\end{table*}

A similar argument may be given for the axial-vector currents in the
Cabibbo model. However, in this case, there are not two well-measured
independent transition matrix elements to be used to define univocally
the rest of the transition matrix elements driven by the weak axial
current.  The only known parameter we have is for the $n\rightarrow p$
weak transition, from where one can extract the axial coupling of the
nucleon, $g_A(0)=g^{np}_1(0)=1.267$. Normally, its $q^2$-dependence is
assumed to have a dipole form with an axial mass of $M_A=1.03$ GeV,
\begin{equation}\label{axial-vector-ff}
 g_A(q^2)=\frac{g_A(0)}{\left(1-
 \frac{q^2}{M^2_A}\right)^2},
\end{equation}
where $g_A(0)=D+F$. One assumption that has been extensively used in
past works \cite{Singh:2006xp,RafiAlam:2019rft,
  Fatima:2018wsy,Fatima:2018tzs, Akbar:2017qsf} assumes that the
$q^2$-dependence acquired by the $D$ and $F$ couplings is identical
and driven by the dependence on $q^2$ of the nucleon axial form factor
$g_A(q^2)$. Under this assumption we can write
\begin{eqnarray}
 &&g^{NY}_1(q^2)=a D_A(q^2)+b F_A(q^2)=
 \frac{a D + b F}{\left(1-
 \frac{q^2}{M^2_A} \right)^2}\nonumber\\
 &=&\frac{a D + b F}{D+F}\frac{g_A(0)}{\left(1-
 \frac{q^2}{M^2_A} \right)^2}=
 \frac{a D + b F}{D+F}g_A(q^2),
\end{eqnarray}
where $a$ and $b$ are SU(3) Clebsch-Gordan coefficients, and $D_A$ and
$F_A$ are normalized to $D$ and $F$ couplings at $q^2=0$. The values
for these axial-vector form factors are tabulated in Table
\ref{tab:axial_ff} for the transitions of interest for our work.

\begin{table*}[ht]
\begin{tabular}{|c|c|c|c|}
\hline
  &  $Y=\Lambda$ & $Y=\Sigma^0$ & $Y=\Sigma^-$\\
 \hline
 $g^{pY}_{1}(q^2)$ &$-\sqrt{\frac16}(1+2x)g_A(q^2)$&
 $\frac{1}{\sqrt{2}}(1-2x)g_A(q^2)$& $0$ \\
 \hline
 $g^{nY}_{1}(q^2)$ & $0$ & $0$ & 
 $(1-2x)g_A(q^2)$\\
 \hline
\end{tabular}
\caption{$g^{NY}_1(q^2)$ axial-vector form factors for the weak
  strangeness-changing transitions considered in this work. The
  definition of $x=\frac{F}{D+F}$ is taken for simplicity in the
  formulae.}\label{tab:axial_ff}
\end{table*}

\begin{figure}[h]
\includegraphics[width=0.35\textwidth,height=.3\textwidth]{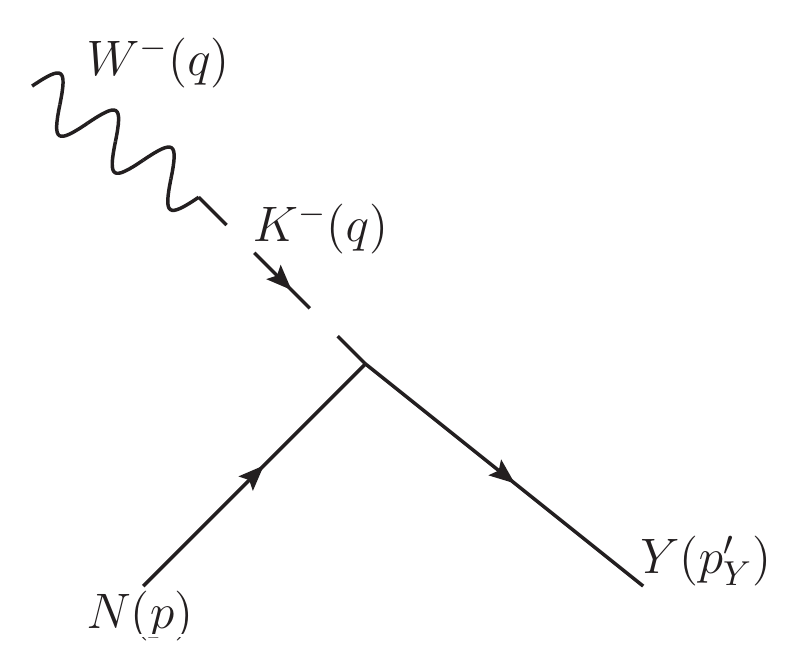}
\caption{Feynman diagram illustrating the generation of the
  pseudo-scalar term in the axial-vector
  current.}\label{fig:kaon_pole}
\end{figure}

Finally, invoking Partial Conservation of the Axial Current (PCAC) in
the chiral limit, we can relate the induced pseudo-scalar
$g^{NY}_3(q^2)$ form factor with the axial one, $g^{NY}_1(q^2)$:
\begin{equation}\label{eq:rel_g3_g1}
g^{NY}_3(q^2)=-g^{NY}_1(q^2)\frac{\left(M+M_Y\right)^2}{q^2}.
\end{equation}
Finally, to take into account the non-vanishing meson masses, the
denominator is extrapolated from $q^2$ to a kaon pole, $q^2-M^2_K$,
for strangeness-changing axial weak charged currents.  This is called
the kaon-pole dominance \cite{Nambu:1960xd}, and it is equivalent to
assume that the induced pseudo-scalar form factor is generated through
the coupling of the $W^{-}$ boson to the baryons through a $K^{-}$, as
depicted in Fig. \ref{fig:kaon_pole}. Although the kaon-pole dominance
is expected to work worse than the pion-pole dominance for non
strangeness-changing weak axial currents, the contribution of the
pseudo-scalar form factor $g^{NY}_3(q^2)$ is proportional to $q^\mu$
and hence to the lepton mass; therefore, its contribution would be too
small for muon and electron antineutrinos induced reactions.

While deriving eq. (\ref{eq:rel_g3_g1}), the baryons in
Fig. \ref{fig:kaon_pole} are taken as on-shell.  The off-shellness of
intermediate baryons can be restored by replacing the $\left(M+
M_Y\right)$ in the numerator with an operator that reduces this factor
when both baryons are on-shell.  That can easily be achieved by
substituting the axial vertex of eq. (\ref{eq:axial_matelem}) by

\begin{widetext}
\begin{equation}
\left\langle Y(p^\prime_Y)\right| A^\mu 
 \left| N(p) \right\rangle = g^{NY}_1(q^2)\;
 \bar{u}_Y(\mathbf{p}^\prime_Y)\left(
  \gamma^\mu\gamma_5 - 
 \frac{q^\mu \slashed{q}}{q^2-M^2_K}\gamma_5
 \right)u_N(\mathbf{p}),
\end{equation}
\end{widetext}
where we used the relationship,
\begin{equation*}
\bar{u}_Y(\mathbf{p}^\prime_Y)\slashed{q}
\gamma_5u_N(\mathbf{p})=\left(M+M_Y\right)
\bar{u}_Y(\mathbf{p}^\prime_Y)\gamma_5u_N(\mathbf{p})
\end{equation*} 
when both baryons are on-shell. 

Now, applying the Feynman rules to the vertices and propagators
appearing in Fig. \ref{fig:background}, which can be extracted from
the Lagrangians given in eqs. (\ref{eq:lag_mesons}) and
(\ref{eq:lag_mb}), we obtain the following hadron currents for the
Born term diagrams:
\begin{widetext}
\begin{eqnarray}
J^\mu_{\rm CT}&=&i\,V_{us}\,\mathcal{A}^{N\rightarrow Y\pi}_{\rm CT}\; 
F_D(q^2)\;\bar{u}_Y(\mathbf{p}_Y) \left[ \gamma^\mu - a^{N\rightarrow Y\pi}
\gamma^\mu \gamma_5 \right] u_N(\mathbf{p})\label{eq:currentCT}\\
J^\mu_{\rm KP}&=&i\,V_{us}\,\mathcal{A}^{N\rightarrow Y\pi}_{\rm KP}\; 
F_D(q^2)\; \frac{q^\mu}{q^2-M^2_K}\; 
\bar{u}_Y(\mathbf{p}_Y) \left[\slashed{q} - 
\frac{\left(M_Y-M\right)}{2} \right]u_N(\mathbf{p})\label{eq:currentKP}\\
J^\mu_{\rm KF}&=&i\,V_{us}\,\mathcal{A}^{N\rightarrow Y\pi}_{\rm KF}\; 
F_D(q^2)\; \frac{2p^\mu_m - q^\mu}{(p_m-q)^2-M^2_K}\; \left(M_Y+M\right)
\bar{u}_Y(\mathbf{p}_Y) \gamma_5 u_N(\mathbf{p})\label{eq:currentKF}\\
J^\mu_{\rm s-Y^\prime}&=&i\,V_{us}\,
\mathcal{A}^{N\rightarrow Y\pi}_{\rm s-Y^\prime}\; 
\bar{u}_Y(\mathbf{p}_Y)\slashed{p}_m\gamma_5 \;
\frac{\slashed{p}+\slashed{q} + M_{Y^\prime}}{(p+q)^2-M^2_{Y^\prime}}\;
\left[ V^\mu_{NY^\prime}(q) - A^\mu_{NY^\prime}(q) \right]u_N(\mathbf{p})
\label{eq:current_sY}\\
J^\mu_{\rm u-N^\prime}&=&i\,V_{us}\,
\mathcal{A}^{N\rightarrow Y\pi}_{\rm u-N^\prime}\; 
\bar{u}_Y(\mathbf{p}_Y)\left[ V^\mu_{N^\prime Y}(q) - 
A^\mu_{N^\prime Y}(q) \right]
\frac{\slashed{p} - \slashed{p}_m + M}{(p-p_m)^2-M^2}\;
\slashed{p}_m\gamma_5 u_N(\mathbf{p})\label{eq:current_uN}
\end{eqnarray},
\end{widetext}
where $Y,Y^\prime=\Sigma,\Lambda$; $N,N^\prime=p,n$; $F_D(q^2)$ is a
global dipole form factor~\footnote{This same assumption for this
global dipole form factor has also been taken in previous works such
as those of Refs.
\cite{RafiAlam:2010kf,Alam:2012zz,Alam:2012ry,Ren:2015bsa}.}
\begin{equation}
F_D(q^2)=\frac{1}{\left(1- \frac{q^2}{M^2_D} \right)^2}, \quad 
M_D\simeq 1\; {\rm GeV}.
\end{equation}
for the CT, KP and KF diagrams.  In
eqs. (\ref{eq:currentCT})-(\ref{eq:current_uN}), the
$\mathcal{A}^{N\rightarrow Y\pi}_i$ are global constants that depend
on the particular reaction given in Table
\ref{tab:constants_diagrams}.

Finally, the vector and axial-vector weak vertices of eqs.
(\ref{eq:current_sY}) and (\ref{eq:current_uN}) are given by
\begin{eqnarray*}
V^\mu_{NY^\prime}(q)&=&f^{NY^\prime}_1(q^2) \gamma^\mu + 
\frac{if^{NY^\prime}_2(q^2)}{M+M_{Y^\prime}} 
\sigma^{\mu\nu} q_{\nu}\\
A^\mu_{NY^\prime}(q)&=&g^{NY^\prime}_1(q^2)\left( 
\gamma^\mu - \frac{q^\mu \slashed{q}}{q^2-M^2_K}\right)\gamma_5,
\end{eqnarray*}
with the vector $f^{NY^\prime}_{1,2}(q^2)$ and axial-vector
$g^{NY^\prime}_1(q^2)$ form factors given in Tables
\ref{tab:vector_ff} and \ref{tab:axial_ff}, respectively.

\begin{table*}[htb]
\begin{tabular}{|c|c|c|c|c|c|c|c|}
\hline
Reaction  & $\mathcal{A}^{N\rightarrow Y\pi}_{\rm CT}$ & 
$a^{N\rightarrow Y\pi}$ & $\mathcal{A}^{N\rightarrow Y\pi}_{\rm KP}$ & 
$\mathcal{A}^{N\rightarrow Y\pi}_{\rm KF}$ & 
$\mathcal{A}^{N\rightarrow Y\pi}_{\rm s-\Sigma}$ & 
$\mathcal{A}^{N\rightarrow Y\pi}_{\rm u-N^\prime}$ & 
$\mathcal{A}^{N\rightarrow Y\pi}_{\rm s-\Lambda}$\\
\hline
$\bar{\nu}_l + p \rightarrow l^{+} + \pi^0 + \Lambda$ & 
$\frac{\sqrt{3}}{2\sqrt{2}f_\pi}$ & $F + \frac{D}{3}$ & 
$-\frac{\sqrt{3}}{2\sqrt{2}f_\pi}$ &$-\frac{(D+3F)}{2\sqrt{6}f_\pi}$ & 
$\frac{D}{\sqrt{3}f_\pi}$ & $\frac{D+F}{2f_\pi}$ & 0\\
\hline
$\bar{\nu}_l + n \rightarrow l^{+} + \pi^{-} + \Lambda$ & 
$\frac{\sqrt{3}}{2f_\pi}$ & $F + \frac{D}{3}$ & 
$-\frac{\sqrt{3}}{2 f_\pi}$ &$-\frac{(D+3F)}{2\sqrt{3}f_\pi}$ & 
$\frac{D}{\sqrt{3}f_\pi}$ & $\frac{D+F}{\sqrt{2}f_\pi}$ & 0\\
\hline
$\bar{\nu}_l + p \rightarrow l^{+} + \pi^{0} + \Sigma^0$ & 
$\frac{1}{2\sqrt{2}f_\pi}$ & $F - D$ & 
$-\frac{1}{2\sqrt{2}f_\pi}$ &$\frac{(D-F)}{2\sqrt{2}f_\pi}$ & 
$0$ & $\frac{D+F}{2f_\pi}$ &$\frac{D}{\sqrt{3}f_\pi}$ \\
\hline
$\bar{\nu}_l + p \rightarrow l^{+} + \pi^{-} + \Sigma^{+}$ & 
$\frac{1}{\sqrt{2}f_\pi}$ & $F - D$ & 
$-\frac{1}{\sqrt{2}f_\pi}$ &$\frac{(D-F)}{\sqrt{2}f_\pi}$ & 
$-\frac{F}{f_\pi}$ & $0$ &$\frac{D}{\sqrt{3}f_\pi}$ \\
\hline
$\bar{\nu}_l + p \rightarrow l^{+} + \pi^{+} + \Sigma^{-}$ & 
$0$ & $0$ & $0$ &$0$ & $\frac{F}{f_\pi}$ & 
$\frac{D+F}{\sqrt{2}f_\pi}$ &$\frac{D}{\sqrt{3}f_\pi}$ \\
\hline
$\bar{\nu}_l + n \rightarrow l^{+} + \pi^{-} + \Sigma^{0}$ & 
$-\frac{1}{2f_\pi}$ & $F-D$ & $\frac{1}{2f_\pi}$ &
$\frac{(F-D)}{2f_\pi}$ & $\frac{F}{f_\pi}$ & 
$\frac{D+F}{\sqrt{2}f_\pi}$ &$0$ \\
\hline
$\bar{\nu}_l + n \rightarrow l^{+} + \pi^{0} + \Sigma^{-}$ & 
$\frac{1}{2f_\pi}$ & $F-D$ & $-\frac{1}{2f_\pi}$ &
$\frac{(D-F)}{2f_\pi}$ & $-\frac{F}{f_\pi}$ & 
$-\frac{D+F}{2f_\pi}$ &$0$ \\
\hline
\end{tabular}
\caption{Constants $\mathcal{A}^{N\rightarrow Y\pi}_i$ and
  $a^{N\rightarrow Y\pi}$ (for the axial-vector piece of the CT
  diagram) for each reaction and diagram in our
  model.}\label{tab:constants_diagrams}
\end{table*}

\begin{figure*}
\begin{subfigure}[h]{0.5\textwidth}
\includegraphics[width=0.95\textwidth,height=.45\textwidth]{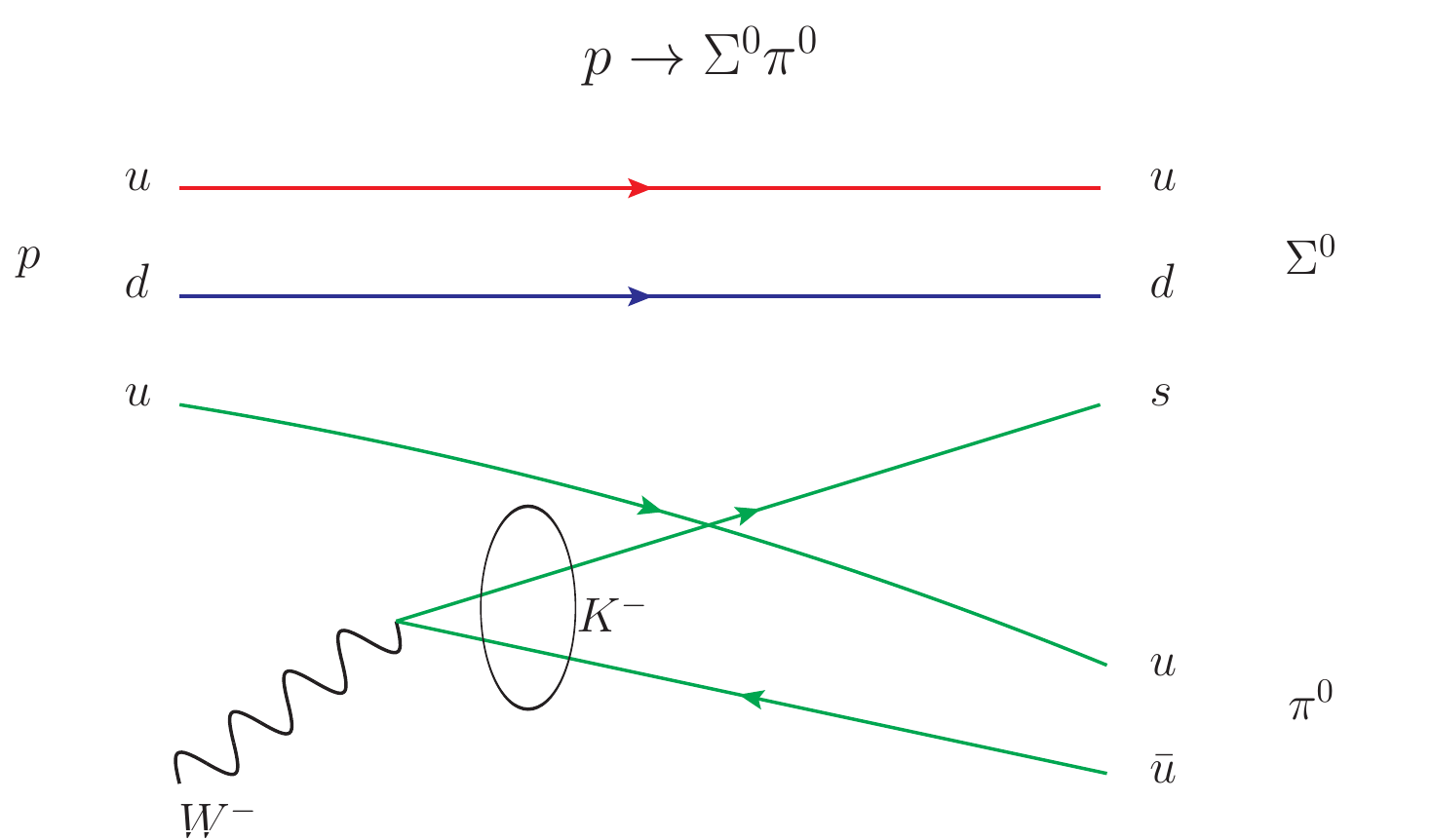}
\caption{In this diagram for the channel $p\rightarrow\Sigma^0\pi^0$,
  the valence quarks of the initial state particles can be fully
  accommodated in the final state particles without any gluon
  emission.}\label{fig:CT_allowed}
\end{subfigure}
\begin{subfigure}[h]{0.49\textwidth}
\includegraphics[width=0.95\textwidth,height=.45\textwidth]{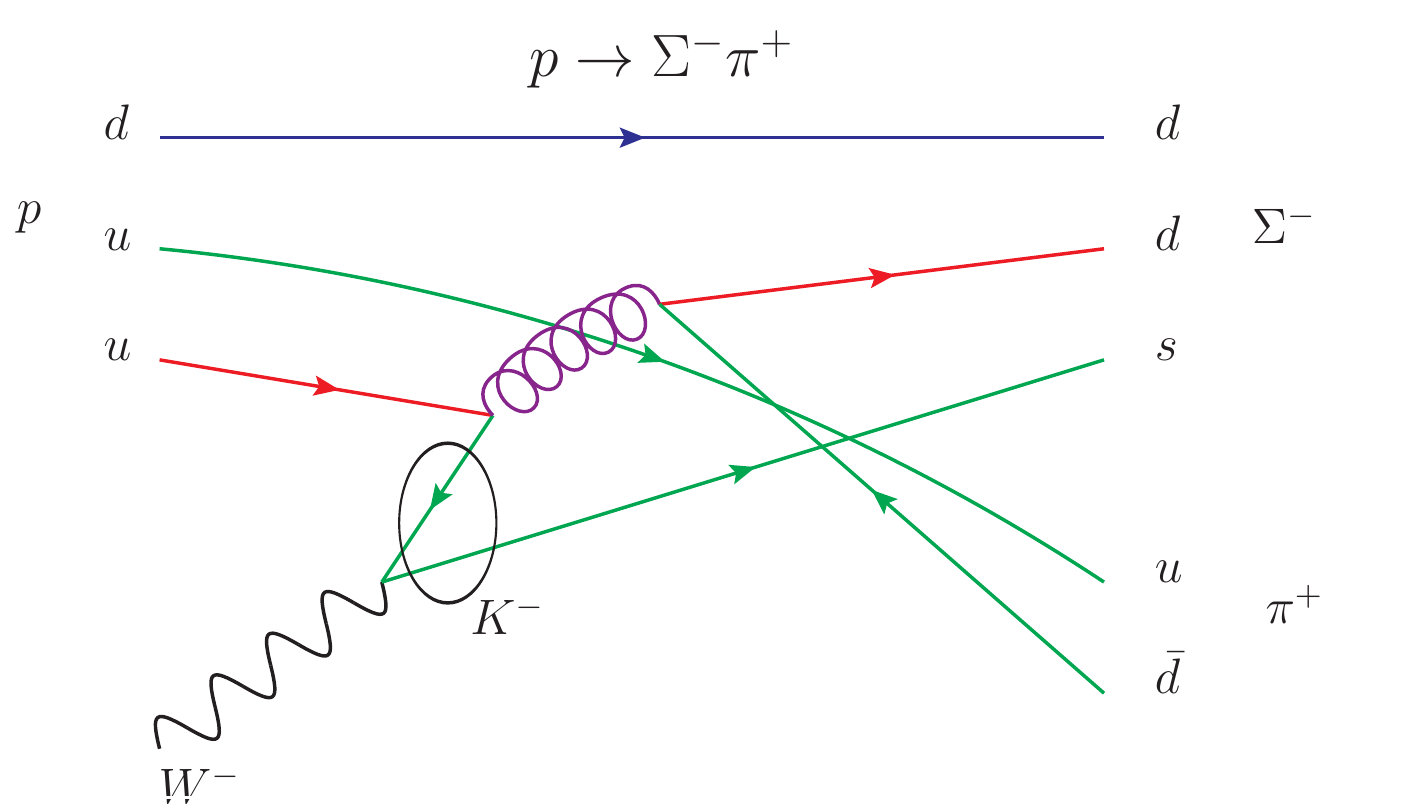}
\caption{In this diagram for the channel
  $p\rightarrow\Sigma^{-}\pi^{+}$, all the valence quarks of the
  initial state particles cannot be fully accommodated in the final
  state particles without gluon emission and the creation of a
  $d\bar{d}$ pair.}\label{fig:CT_forbidden}
\end{subfigure}
\caption{Two possible Feynman diagrams in terms of quarks and gluons
  to explain why the CT and KP diagrams are forbidden for the
  $p\rightarrow\Sigma^{-}\pi^{+}$ reaction channel but not for the
  others.  The colored quark lines represent their possible colors in
  QCD to make colorless initial and final hadrons.}
\label{fig: CT}
\end{figure*}

The fact that the CT and KP diagrams for the
$p\rightarrow\Sigma^{-}\pi^{+}$ channel are zero and not for the other
ones can be explained with the help of Figs. \ref{fig:CT_allowed} and
\ref{fig:CT_forbidden}. The key is not to need to emit gluons in these
diagrams, i.e, that the virtual $s\bar{u}$ pair ($K^{-}$) in which the
$W^{-}$ decays could be redistributed along with the quarks of the
initial nucleon in the two final hadrons, the hyperon and the pion,
but without the need of emitting gluons to create a $q\bar{q}$ pair of
the same flavor. It seems to be a kind of OZI forbidding rule because
the valence quarks of the initial $W^{-}N$ state get fully
redistributed into the final $Y\pi$ state without any gluon
emission. This is totally possible for all the channels except for the
$W^{-}p\rightarrow\Sigma^{-}\pi^{+}$ as shown in
Fig. \ref{fig:CT_forbidden}. Notice that the $\bar{u}$ antiquark
coming out from the decay of the $W^{-}$ is not present in the final
state. Therefore, it is completely necessary to annihilate it with an
$u$ quark via gluon emission to have the right quarks in the final
state.

As the $s\bar{u}$ quark-antiquark pair has the same quantum numbers as
the $K^{-}$, this argument holds not only for the CT diagram, but for
the KP as well.

It is worth noting that these findings are in agreement for the tree
level amplitudes for the reaction channel
$p\rightarrow\Sigma^{-}\pi^{+}$ with those of Ref. \cite{Ren:2015bsa},
where the authors consider the CT, KP and KF reaction mechanisms.

This argument also explains why in Ref. \cite{RafiAlam:2019rft} there
were no CT and KP amplitudes at tree level.

\subsection{Resonance model}\label{subsec:resonance}
To describe the currents of the resonance diagrams depicted in
Fig. \ref{fig:resonances}, we follow the prescription discussed in
Refs. \cite{Hernandez:2007qq,
  Alam:2015gaa,Alam:2012zz,Alam:2012ry,RafiAlam:2019rft} and include
the lowest lying resonances belonging to the decuplet representation
of the SU(3) group. The resonant states which may appear in the
s-channel and u-channel are $\Sigma^*(1385)$ and $\Delta(1232)$,
respectively.

Though the $\Delta(1232)$ resonances are widely studied in the
literature, there is less information available for the
$\Sigma^*(1385)$ resonances.  However, we know that both
$\Sigma^*(1385)$ and $\Delta(1232)$ are members of the same decuplet,
therefore under the assumption of exact SU(3) flavor symmetry for the
couplings and using the eq. (\ref{eq:CG_series}), the weak transition
form factors connecting an octet state to a decuplet state can be
obtained.  One should notice that as the weak charged current belongs
to the octet representation of current operators of the SU(3) group,
and couples one octet state with a decuplet state, the representation
$\left\lbrace 10 \right\rbrace$ appears only once in the
Clebsch-Gordan series of eq. (\ref{eq:CG_series}). Therefore, there is
only one independent reduced matrix element. We will take for the
latter the transition matrix element as:
\begin{equation}\label{eq:nDelta_matelem}
\left\langle \Delta^{+}(p_R) \right| j^\mu_{\Delta S=0} 
\left| n(p) \right\rangle = \bar{u}_{\alpha}(\mathbf{p}_R) 
\Gamma^{\alpha\mu}(p,q) u(\mathbf{p}),
\end{equation} 
with $p_R=p+q$. In eq. (\ref{eq:nDelta_matelem}),
$\Gamma^{\alpha\mu}(p,q)$ is the vertex function given by
\begin{widetext}
\begin{eqnarray}
\Gamma^{\alpha\mu}(p,q)&=&\Bigg[\frac{C^V_3}{M}\left(
g^{\alpha\mu}\slashed{q}-q^{\alpha}\gamma^{\mu}\right) 
+ \frac{C^V_4}{M^2}\left(g^{\alpha\mu}q\cdot(p+q)-
q^\alpha(p+q)^\mu\right)
+ \frac{C^V_5}{M^2}\left( g^{\alpha\mu} q\cdot p - 
q^{\alpha}p^\mu\right)+C^V_6 g^{\alpha\mu}\Bigg]\gamma_5\nonumber\\
&+& \Bigg[ \frac{C^A_3}{M}\left(
g^{\alpha\mu}\slashed{q}-q^{\alpha}\gamma^{\mu}\right) 
+\frac{C^A_4}{M^2}\left(g^{\alpha\mu}q\cdot(p+q)-
q^\alpha(p+q)^\mu\right) 
+C^A_5 g^{\alpha\mu} + \frac{C^A_6}{M^2}q^\alpha q^\mu \Bigg],
\label{eq:vertex_gamma}
\end{eqnarray}
\end{widetext}
$\bar{u}_{\alpha}(\mathbf{p}_R)$ is a Rarita-Schwinger spinor
describing spin-$\frac32$ particles, and $j^\mu_{\Delta S=0}$ is the
strangeness-preserving weak charged current coupled to an incoming
$W^{+}$ boson.

A systematic way of obtaining the relationships (SU(3) factors)
between the weak vertices for all the allowed transitions and that for
the $n\rightarrow\Delta^+$ (given in eq. (\ref{eq:vertex_gamma})) is
to use the lowest order Lagrangian that couples the decuplet baryons
with the octet baryons and mesons in the presence of an external
current \cite{Butler:1992pn,Doring:2006ub} and that was already used
in Refs. \cite{Alam:2012zz,Alam:2012ry, RafiAlam:2019rft}. Its form is
\begin{equation}\label{eq:lag_dec}
\mathcal{L}_{\rm dec}= \mathcal{C}
\left( \epsilon^{abc}\,
\overline{T}^\mu_{ade} (u_{\mu})^{d}_{b}\, B^{e}_{c} 
+\epsilon^{abc} \bar{B}^{c}_{e} (u_\mu)^{b}_{d}\, T^{\mu}_{aed}\right),
\end{equation}
where $B$ is given by eq. (\ref{eq:baryon_matrix}), $u_\mu$ is the
vielbein of eq. (\ref{eq:vielbein}), and $T^\mu_{aed}$ is the SU(3)
representation of the Rarita-Schwinger fields for the decuplet
baryons.  This representation is completely symmetric in the three
flavor indices, and an implicit sum over flavor indices
($a,b,...=1,2,3$) is understood in eq.  (\ref{eq:lag_dec}). It is
worth relating the $T_{abc}$ representation to the physical
states~\footnote{Note that there is a typographical mistake in
$T_{233}$ for the $\Xi^{*-}$ state in the footnotes of
Refs. \cite{Alam:2012zz,Alam:2012ry}.}:
\begin{eqnarray}
&&T_{111}= \Delta^{++}; \quad T_{112}=\frac{\Delta^{+}}{\sqrt{3}};
\quad T_{122}=\frac{\Delta^{0}}{\sqrt{3}}\nonumber\\
&& T_{222}=\Delta^{-}; \quad T_{113}=\frac{\Sigma^{*+}}{\sqrt{3}};
\quad T_{123}=\frac{\Sigma^{*0}}{\sqrt{6}}\nonumber\\
&& T_{223}=\frac{\Sigma^{*-}}{\sqrt{3}}; \quad 
T_{133}=\frac{\Xi^{*0}}{\sqrt{3}}; \quad 
T_{233}=\frac{\Xi^{*-}}{\sqrt{3}}\nonumber\\
&&T_{333}=\Omega^{-}.
\end{eqnarray}

The Lagrangian of eq. (\ref{eq:lag_dec}) only provides the leading
weak axial coupling $C^A_5(0)$ for all the allowed weak transitions.
Knowing that $C^A_5(0)|_{n\rightarrow\Delta^{+}}
\simeq\frac{2\mathcal{C}}{\sqrt{3}}$ with
$\mathcal{C}\sim1$~\cite{Alam:2012zz,RafiAlam:2019rft}, we can relate
all the other leading axial couplings for the other weak transitions
to the $n\rightarrow\Delta^{+}$ one. These relative factors are then
applied to all the vector $C^V_i(q^2)$ and axial $C^A_i(q^2)$ form
factors, thus assuming exact SU(3) symmetry for the
couplings~\footnote{In Appendix~\ref{sec:appendix_a}, we give an
equivalent formulation based on flavor SU(3) symmetry.}.  We choose
the form factors for the $n \rightarrow \Delta^{+}$ transition given
in Ref.  \cite{Hernandez:2007qq} with the exception that for
strangeness-changing processes
\begin{equation}
C^A_6(q^2)=C^A_5(q^2)\frac{M^2}{M^2_K-q^2},
\end{equation}
which appears when one imposes PCAC for the transition similar to
Fig. \ref{fig:kaon_pole} with the final hyperon replaced by the
$\Sigma^*(1385)$ resonance.  Note that the strong coupling
$\mathcal{C}\simeq 1$ is obtained to match the $\Delta$ width at its
nominal mass.  If we apply the Feynman rules to the diagrams depicted
in Fig. \ref{fig:resonances}, we obtain the following amplitudes:
\begin{widetext}
 \begin{eqnarray}
  J^{\mu}_{\rm s-\Sigma^*}&=&
  i\,V_{us}\, 
  \mathcal{A}^{N\rightarrow Y\pi}_{\rm s-\Sigma^*}\,\frac{p^\beta_m}{p^2_{\Sigma^*}-
  M^2_{\Sigma^*}+iM_{\Sigma^*}
  \Gamma_{\Sigma^*}}\;\bar{u}_Y(\mathbf{p}_Y)\,
  \mathcal{P}_{\beta\alpha}(p_{\Sigma^*})\,
  \Gamma^{\alpha\mu}(p,q)\, u_N(\mathbf{p})
  \label{eq:current_ssigmastar}\\
  J^{\mu}_{\rm u-\Delta}&=& i\, V_{us}\,
  \mathcal{A}^{N\rightarrow Y\pi}_{\rm u-\Delta}\,\frac{p^\beta_m}{p^2_\Delta-M^2_\Delta + i M_\Delta \Gamma_\Delta}\;
  \bar{u}_Y(\mathbf{p}_Y)\,
  \tilde{\Gamma}^{\mu\alpha}(p_Y,q)\,
  \mathcal{P}_{\alpha\beta}(p_\Delta)\,
  u_N(\mathbf{p}),\label{eq:current_delta}
 \end{eqnarray}
\end{widetext}
where $p_{\Sigma^*}=p+q$, $p_\Delta=p-p_m$,
$\tilde{\Gamma}^{\mu\alpha}(p_Y,q)=\gamma^0
\left[\Gamma^{\alpha\mu}(p_Y,-q)\right]^\dagger \gamma^0$,
$\mathcal{P}_{\alpha\beta}(p_D)$ is the spin-$\frac32$ projector
operator appearing in the propagator of Rarita-Schwinger fields, and
given by
\begin{eqnarray}
 \mathcal{P}_{\alpha\beta}(P)&=&
 -\left(\slashed{P} + M_D\right)
 \Bigg[ g_{\alpha\beta} - \frac13 
 \gamma_\alpha \gamma_\beta \nonumber\\
 &-& \frac23\,
 \frac{P_{\alpha}\;P_{\beta}}{M^2_D}
 + \frac13\, \frac{P_{\alpha}\,
 \gamma_\beta - P_{\beta}\,
 \gamma_\alpha}{M_D}\Bigg],\nonumber\\
\end{eqnarray}
with $M_D$ the corresponding mass of the decuplet baryon, either the
$\Delta$ or the $\Sigma^*$, and $P$ the four-momentum carried by these
particles.  The constants $\mathcal{A}^{N\rightarrow Y\pi}_i$
appearing in eqs. (\ref{eq:current_ssigmastar}) and
(\ref{eq:current_delta}) are given in table
\ref{tab:constants_diagrams_resonances}.

\begin{table*}[htb]
\begin{tabular}{|c|c|c|}
\hline
Reaction  & $\mathcal{A}^{N\rightarrow Y\pi}_{\rm s-\Sigma^*}$ & 
$\mathcal{A}^{N\rightarrow Y\pi}_{\rm u-\Delta}$\\
\hline
$\bar{\nu}_l + p \rightarrow l^{+} + \pi^0 + \Lambda$ & 
$\frac{\mathcal{C}}{\sqrt{2}f_\pi}$ & $0$\\
\hline
$\bar{\nu}_l + n \rightarrow l^{+} + \pi^{-} + \Lambda$ & 
$\frac{\mathcal{C}}{f_\pi}$ & $0$\\
\hline
$\bar{\nu}_l + p \rightarrow l^{+} + \pi^{0} + \Sigma^0$ & 
$0$ & $2\sqrt{\frac23}\frac{\mathcal{C}}{f_\pi}$\\
\hline
$\bar{\nu}_l + p \rightarrow l^{+} + \pi^{-} + \Sigma^{+}$ & 
$\frac{\mathcal{C}}{\sqrt{6}f_\pi}$ & $\frac{\mathcal{C}\sqrt{6}}{f_\pi}$\\
\hline
$\bar{\nu}_l + p \rightarrow l^{+} + \pi^{+} + \Sigma^{-}$ & 
$-\frac{\mathcal{C}}{\sqrt{6}f_\pi}$ & $\sqrt{\frac23}\frac{\mathcal{C}}{f_\pi}$\\
\hline
$\bar{\nu}_l + n \rightarrow l^{+} + \pi^{-} + \Sigma^{0}$ & 
$-\frac{\mathcal{C}}{\sqrt{3}f_\pi}$ & $-\frac{2\mathcal{C}}{\sqrt{3}f_\pi}$\\
\hline
$\bar{\nu}_l + n \rightarrow l^{+} + \pi^{0} + \Sigma^{-}$ & 
$\frac{\mathcal{C}}{\sqrt{3}f_\pi}$ & $\frac{2\mathcal{C}}{\sqrt{3}f_\pi}$\\
\hline
\end{tabular}
\caption{Constants $\mathcal{A}^{N\rightarrow Y\pi}_i$ for each
  reaction and the resonances (s-$\Sigma^*$ and u-$\Delta$) diagrams
  of Fig. \ref{fig:resonances} in our model.}
\label{tab:constants_diagrams_resonances}
\end{table*}

Finally, in eq. (\ref{eq:current_ssigmastar}), $\Gamma_{\Sigma^*}$ is
the energy dependent $\Sigma^*(1385)$ width, given by
\begin{equation*}
\Gamma_{\Sigma^*}=\Gamma_{\Lambda\pi}+\Gamma_{\Sigma\pi}+
\Gamma_{N\bar{K}} + \Gamma_{\Sigma\eta} + \Gamma_{\Xi K},
\end{equation*}
where the different strong partial widths $\Gamma_{B\phi}$ can be
calculated with the vertices from the Lagrangian given in
eq. (\ref{eq:lag_dec}). Their expressions are always the same up to an
SU(3) factor and are given as
\begin{eqnarray}
\Gamma_{D\rightarrow B\phi}&=& \frac{C_{B\phi}}{192\pi}
\left(\frac{\mathcal{C}}{f_\pi}\right)^2 \frac{(W+M_B)^2-m^2_\phi}{W^5}
\nonumber\\
&&\lambda^{3/2}(W^2,M^2_B,m^2_\phi)\; \Theta(W-M_B-m_\phi),\nonumber\\
\end{eqnarray}
where $\lambda(x,y,z)=x^2+y^2+z^2-2xy-2xz-2yz$ is the K\"allen
$\lambda$-function, $M_B$ and $m_\phi$ are the final baryon and meson
masses in the decay of the $\Sigma^*$, $\Theta$ is the unit step
function allowing a partial width to decay into channels $B\phi$, only
when the invariant mass squared $W^2=(p+q)^2$ carried by the resonance
is higher than the threshold $(M_B+m_\phi)^2$.  Finally, the SU(3)
factors $C_{B\phi}$ are 1 for $\Lambda\pi$ and $\Sigma\eta$, while
they are $\frac23$ for the $\Sigma\pi$, $N\bar{K}$ and $\Xi K$.

In eq. (\ref{eq:current_delta}), for the $\Delta(1232)$ appearing in
u-channel diagram we take $\Gamma_\Delta \to 0$ for the present
kinematics, as the four momentum $p_{\Delta}(=p-p_m)$ squared is
always below the decay threshold. Indeed,
 \begin{eqnarray}
 p^2_{\Delta}&=&M^2+m^2_\pi-2ME_\pi \leqslant 
 (M-m_\pi)^2 < (M+m_\pi)^2\nonumber\\
 &<& M^2_{\Delta}. 
 \label{eq:pdelta2}
\end{eqnarray}
This leads to the $\Delta$ width equals zero as $p^2_\Delta <
(M+m_\pi)^2$ holds for all the kinematics regions under consideration.

\section{Results}\label{sec:results}
\subsection{Total cross sections}
\label{subsec:total_xsect}

\begin{figure*}
\includegraphics[width=0.45\textwidth,height=.4\textwidth]{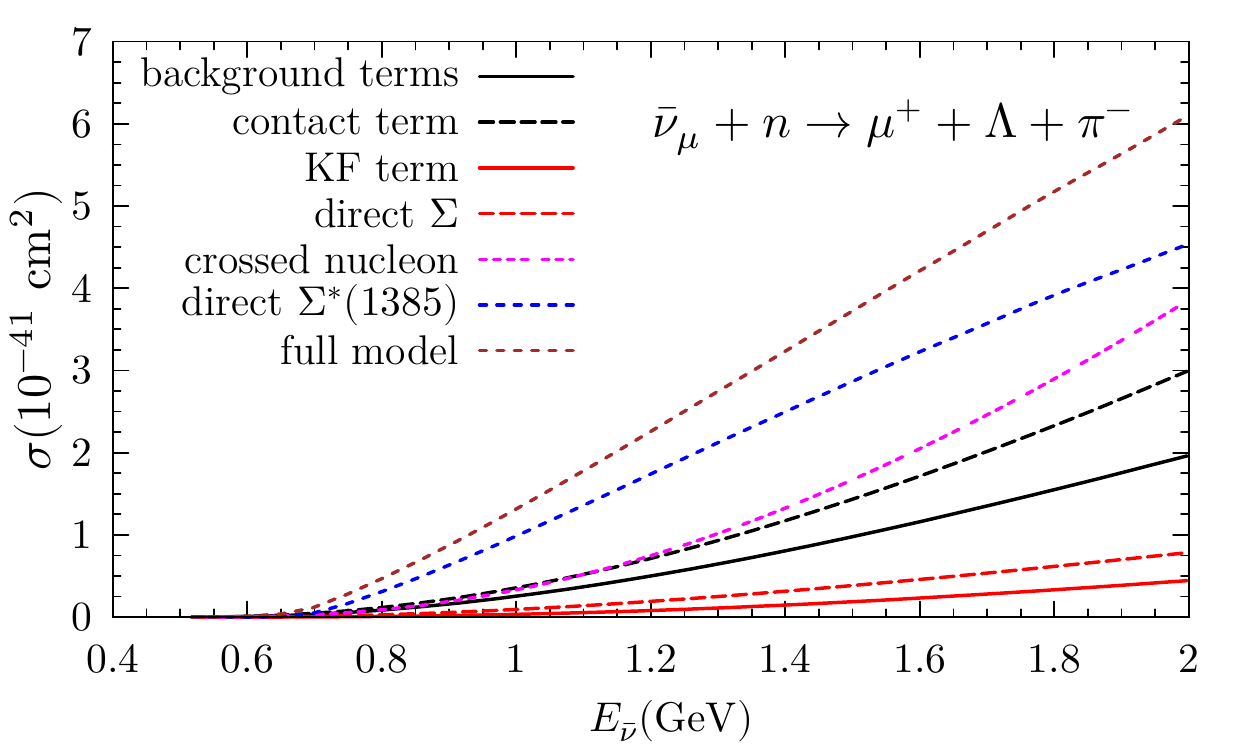}
\includegraphics[width=0.45\textwidth,height=.4 \textwidth]{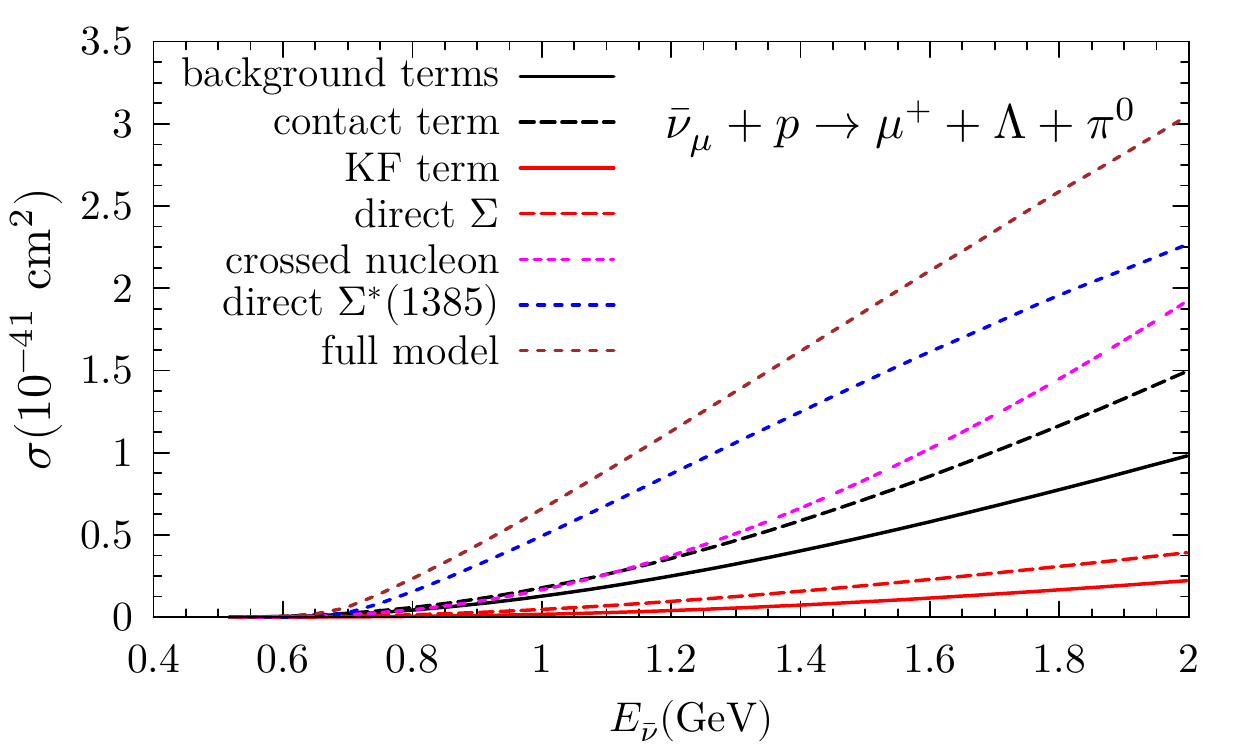}
\caption{Total cross sections for the $\Lambda$ hyperon production off
  neutrons (left panel) and protons (right panel).  Some of the
  contributions of individual diagrams of Figs. \ref{fig:background}
  and \ref{fig:resonances} have been singled out. Note that the nature
  is identical in both panels, except the scale in the vertical axis.
  This is because the total cross section for neutrons is exactly
  twice that for protons (see appendix \ref{sec:appendix_a}).}
\label{fig:lambda_xsect}
\end{figure*}

\begin{figure}
\includegraphics[width=0.45\textwidth,height=.4\textwidth]{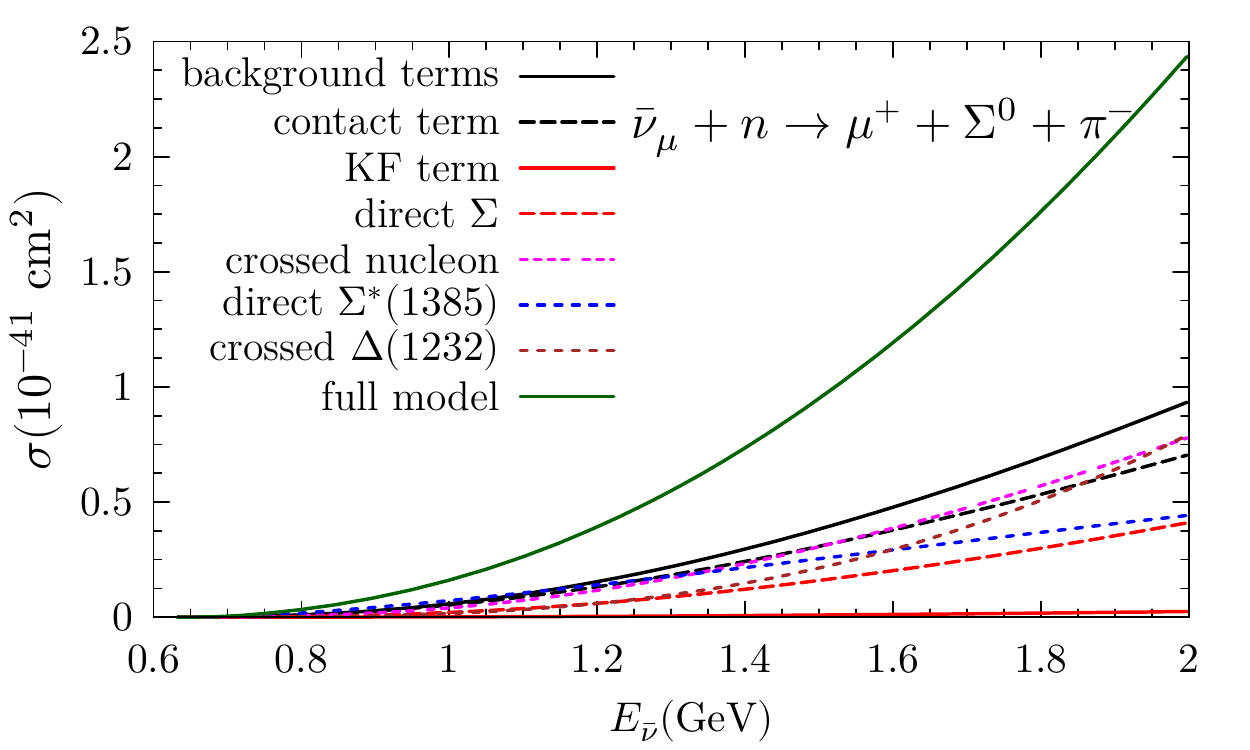}
\caption{Total cross sections for the $\Sigma^0\pi^{-}$ and
  $\Sigma^{-}\pi^0$ production off neutrons. We present here results
  for $\Sigma^0\pi^{-}$ production only. The results for the
  $\Sigma^{-}\pi^0$ are identical as the hadron amplitude is the same
  up to a relative sign (see appendix \ref{sec:appendix_a}). We also
  present individual contributions of some of the diagrams following
  Fig.~\ref{fig:lambda_xsect}.}
\label{fig:neutron_sigma_xsect}
\end{figure}

\begin{figure*}
\includegraphics[width=0.32\textwidth,height=.34 \textwidth]{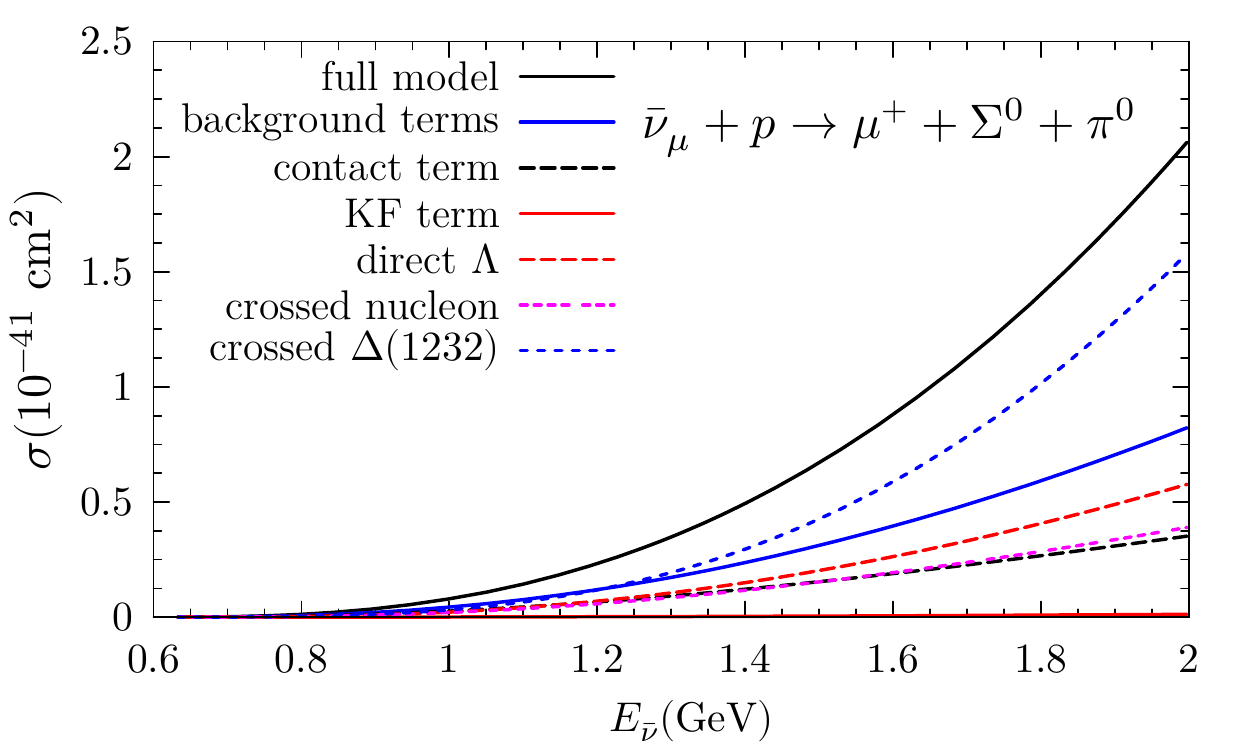}
\includegraphics[width=0.32\textwidth,height=.34 \textwidth]{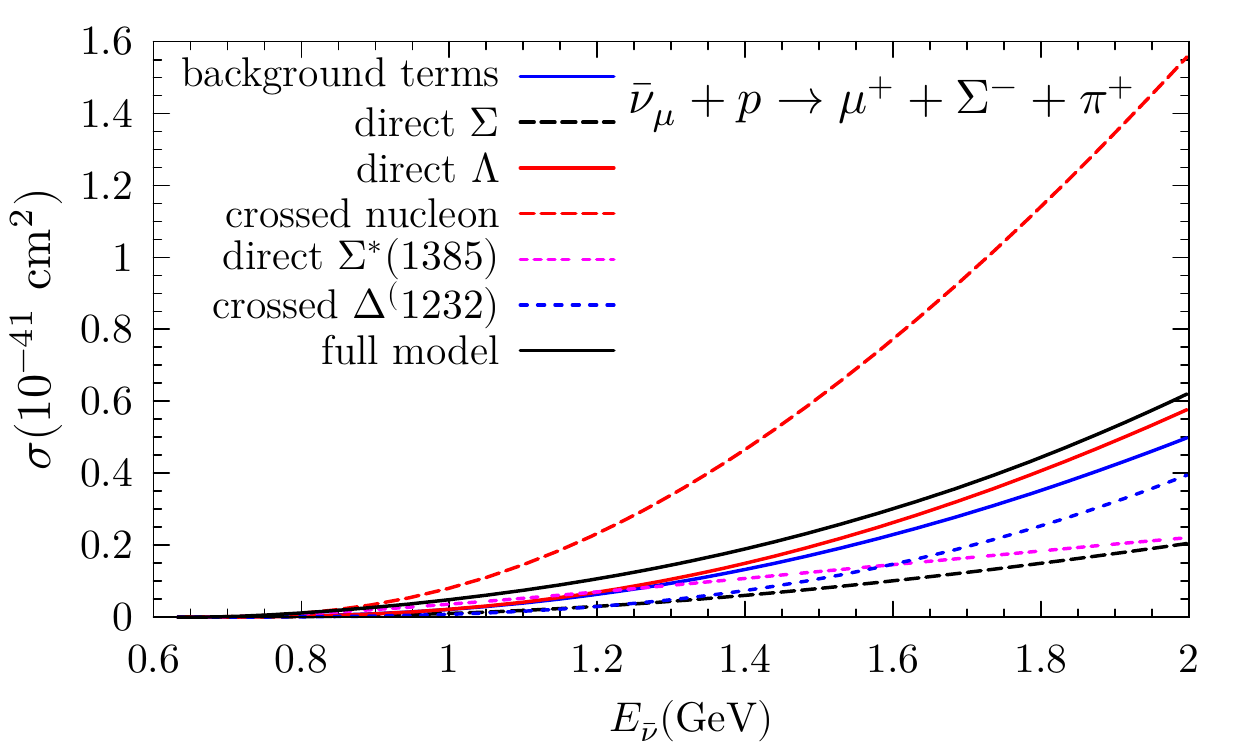}
\includegraphics[width=0.32\textwidth,height=.34 \textwidth]{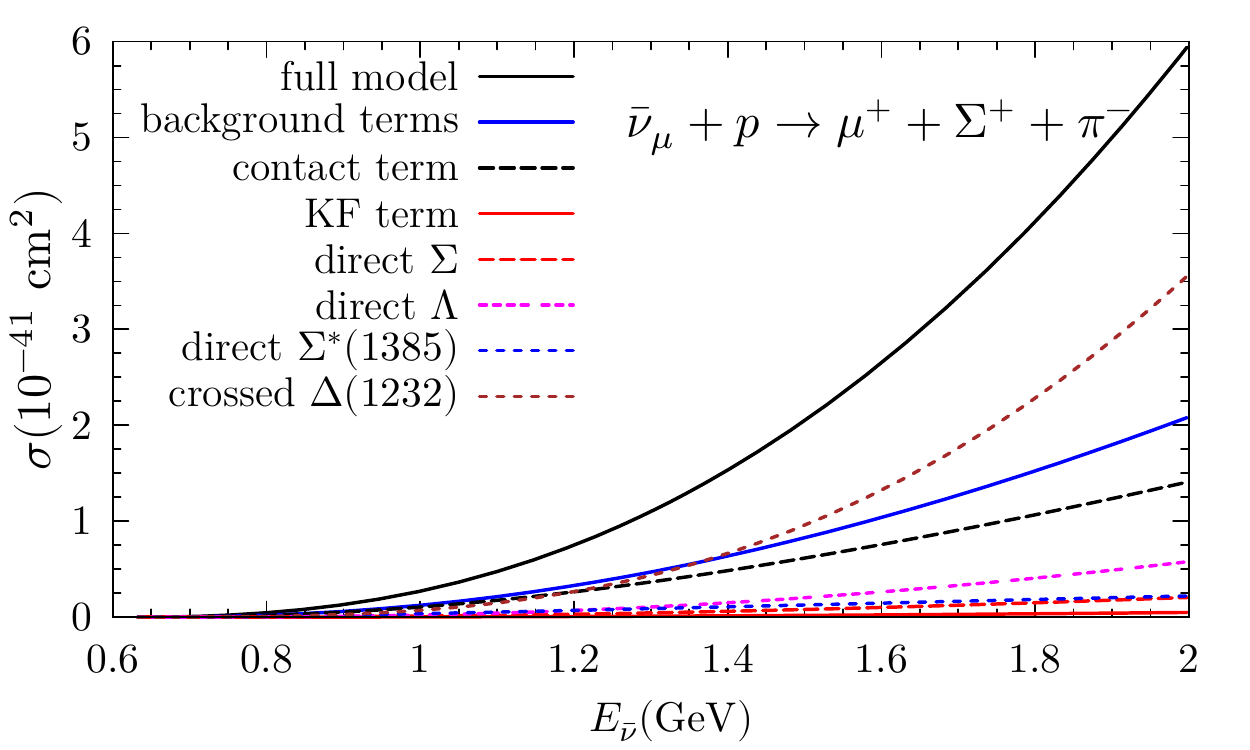}
\caption{Total cross sections for the $\Sigma$ hyperon production off
  protons, $\Sigma^0\pi^{0}$ on the left panel, $\Sigma^{-}\pi^{+}$ on
  the middle one and $\Sigma^{+}\pi^{-}$ on the right.  The individual
  contribution is also shown, similar to Fig.~\ref{fig:lambda_xsect}.}
\label{fig:proton_sigma_xsect}
\end{figure*}

In Figs. \ref{fig:lambda_xsect}, \ref{fig:neutron_sigma_xsect} and
\ref{fig:proton_sigma_xsect} we show results for the total cross
sections off proton and neutron targets, as a function of the
muon-antineutrino energy in the LAB frame.  In order to understand the
dynamics of the reaction channels, we show the contribution due to
individual diagrams of Figs~\ref{fig:background} and
\ref{fig:resonances}, wherever applicable. One may notice that we do
not give the results for all the channels due to the KP diagram of
Fig.~\ref{fig:background} . This is because the hadron tensor
associated with the KP diagram alone is proportional to $q^\mu q^\nu$,
and when contracted with the lepton tensor will be proportional to the
square of lepton mass, making their individual contributions
negligible for electron and muon antineutrinos.  However, they are
present in ``full-model". A detailed discussion over the individual
contributions of all the diagrams will be given later in this section.

The $\Lambda\pi$ final state on neutron and proton target are shown in
Fig.~\ref{fig:lambda_xsect}. Apart from the individual contributions,
we also present results for the background terms, where we add all the
Feynman amplitudes of Fig.~\ref{fig:background} coherently. We find
that the background terms are comparable with the resonance
contribution.  One particular feature for the $\Lambda\pi$ production
is that the cross section off the neutron target is exactly twice as
that for proton target; see appendix~\ref{sec:appendix_a} for the
SU(3) relationships derived for the different amplitudes (hadronic
currents).

A similar trend can be found for $\Sigma\pi$-production cross sections
off neutrons, as depicted in Fig~\ref{fig:neutron_sigma_xsect}. In
this case, the possible $\Sigma\pi$-final states are $\Sigma^0 \pi^-$
and $\Sigma^- \pi^0$.  However, it turns out that the cross section
for both channels is exactly identical and is shown for one of the
channels ($\bar \nu_\mu + n \to \mu^+ + \Sigma^0 + \pi^-$). This can
be understood from their isospin relations as given in appendix
\ref{sec:appendix_a}, where the modulus of the isospin factors are the
same for both the channels.  While the individual diagrams contribute
similarly to $\Lambda\pi$ production, the full model grows faster for
the $\Sigma\pi$ reaction.

The relative size of the contributions of many mechanisms depicted in
Figs.  \ref{fig:background} and \ref{fig:resonances} can be understood
in terms of their couplings alone, given by the constants
$\mathcal{A}^{N\rightarrow Y\pi}_i$ of tables
\ref{tab:constants_diagrams} and
\ref{tab:constants_diagrams_resonances}.  For instance, the smallness
of the KF contributions in the reaction channels producing $\Sigma$
hyperons (Figs. \ref{fig:neutron_sigma_xsect} and
\ref{fig:proton_sigma_xsect}) can be explained because their cross
sections are proportional to the square of $(D-F)$, while for the
reactions producing $\Lambda$ hyperons, these are proportional to the
square of $(D+3F)$, which is much larger.  Further, if we compare
$\Lambda \pi$ with $\Sigma \pi$ final states close to threshold
energies, $\Lambda \pi$ cross section is higher than that of $\Sigma
\pi$ as $M_\Lambda < M_\Sigma$, thus allowing a larger phase space for
the same antineutrino energies. However, the overall contribution of
KF diagram is relatively low, as the virtual $K$ in the KF diagram is
carrying a four-momentum which is highly off-shell,
\begin{equation}\label{eq:pk2_offshell}
p^2_K=(p-p_Y)^2\leqslant(M-M_Y)^2\ll M^2_K, 
\end{equation}
which also suppresses its contribution.

In fact, if one looks at the studies carried out in Refs.
\cite{RafiAlam:2010kf,Alam:2012zz,Alam:2012ry}, this kind of
contributions is more sizeable when the mass of the exchanged meson is
lighter, as it is the case of the $\pi P$ diagrams with respect to the
$\eta P$ ones, if one inspects some of the figures depicted in
Refs. \cite{RafiAlam:2010kf,Alam:2012zz,Alam:2012ry}.

Next, we explore the crossed-nucleon diagrams.  In general, the
crossed-nucleon diagrams are important because of two main reasons:
the constants of the diagrams $\mathcal{A}^{N\rightarrow Y\pi}_{\rm
  u-N^\prime}$ are proportional to $(D+F)$ coupling coming from the
$NN^\prime\pi$ vertex (see table \ref{tab:constants_diagrams}), which
is also large; and because the four-momentum squared carried by the
intermediate nucleon is closer to its squared mass, given that the
mass of the final $\pi$ meson is light (see eq. (\ref{eq:pdelta2})
with $M^2_\Delta$ replaced by $M^2$).  Therefore, in this case, the
difference in the intermediate nucleon propagator, $(p-p_m)^2-M^2$, is
smaller in absolute value than for the crossed-$\Delta$ propagator.
The relative size of the crossed-diagrams for the different channels
can be understood using table~\ref{tab:constants_diagrams} along with
tables~\ref{tab:vector_ff} and \ref{tab:axial_ff}.  For example, in
Fig.~\ref{fig:proton_sigma_xsect}, the ratio
$\sigma_{u-N^\prime}(\Sigma^0 \pi^0) : \sigma_{u-N^\prime}(\Sigma^-
\pi^+) $ is $1:4$, due to the square of constants $\mathcal{A}_i$ and
the square of the ratio of vector and axial-vector transition form
factors for $p\rightarrow\Sigma^0$ and $n\rightarrow\Sigma^{-}$ in
tables \ref{tab:vector_ff} and \ref{tab:axial_ff}.  Something similar
happens with the neutron induced $\Sigma\pi$ reactions of
Fig. \ref{fig:neutron_sigma_xsect}, but in this case the factors
compensate, giving the same contribution ($1:1$ ratio) to the cross
section.
 
 In the $s$-channel we find that, normally, the direct $\Lambda$
 contributions are larger than the direct $\Sigma$ ones off protons by
 a factor $\sim3$ when both diagrams are present in the same reaction
 channel. This can be more or less understood because
 $\frac{D}{\sqrt{3}}\sim F$ and if one neglects (which is not a bad
 approximation for the vector form factors) the contribution of the
 charge $f^n_1(q^2)$ (certainly not good for the magnetic
 $f^n_2(q^2)$) form factor, the ratio of direct $\Lambda$ over direct
 $\Sigma$ is roughly $\left(\frac{D}{F}\right)^2\sim3$.  Of course,
 the pure axial-vector contribution and the interference vector-axial
 in those diagrams seem to have a trend to cancel because otherwise,
 the ratio $3:1$ would not be so accurate as it happens to be.
 
The contribution of direct $\Sigma^*$ resonance channel is very
important for the final $\Lambda\pi$ production reactions of
Fig.~\ref{fig:lambda_xsect}, and gradually decreases for $\Sigma\pi$
production off neutrons (Fig. \ref{fig:neutron_sigma_xsect}) and off
protons (Fig. \ref{fig:proton_sigma_xsect}). The reason is two-fold:
on the one hand, the ratio $\mathcal{A}_{\rm
  s-\Sigma^*}^{n\rightarrow\Sigma\pi} \; : \; \mathcal{A}_{\rm
  s-\Sigma^*}^{n\rightarrow\Lambda\pi^{-}} $ is $1: \sqrt{3}$, and
hence a factor $3$ of reduction in the cross section is obtained off
the neutron channels; similarly, for the reaction off protons a factor
$6$ of reduction in the cross section is found. Additionally, on the
other hand, at low energies, the $\Lambda\pi$ production channel
dominates over $\Sigma\pi$ because of the threshold effect. This
threshold effect, together with smaller couplings for $\Sigma\pi$
channels, reduces the contribution of the $\Sigma^*$ resonance for the
final production of $\Sigma\pi$.
 
On the other hand, the crossed $\Delta$ diagrams are important for the
$\Sigma\pi$ reaction channels, especially when induced off protons
(see Fig. \ref{fig:proton_sigma_xsect}).  In fact, from table
\ref{tab:constants_diagrams_resonances} the cross sections for the
channels $p\rightarrow\Sigma^{+}\pi^{-}$,
$p\rightarrow\Sigma^{0}\pi^{0}$, $n\rightarrow\Sigma\pi$ and
$p\rightarrow\Sigma^{-}\pi^{+}$ are found in the relative ratios
$9:4:2:1$, respectively.
 
In general, the interferences between the different mechanisms
(diagrams) are significant and destructive, except for the $ p \to
\Sigma^+ \pi^-$ channel, see Fig.~\ref{fig:proton_sigma_xsect}. For
all other channels under consideration, we find that the interferences
are important and reduce the total cross section compared with the
incoherent sum of the singled-out contributions. In some cases, like
in the reaction $\bar{\nu}_{\mu}+p\rightarrow \mu^{+} +
\Sigma^{-}+\pi^{+}$ (see Fig.~\ref{fig:proton_sigma_xsect}) a single
mechanism is much larger than the total cross section.  Similar
results are found for the $\Lambda\pi$ production, as might be seen in
Fig. \ref{fig:lambda_xsect}. Here we must point out that the chiral
Lagrangian fixes the relative sign between all (non-resonant)
diagrams, at least close to the threshold.
 
 \begin{figure}[htb]
  \includegraphics[width=0.45\textwidth,height=.4 \textwidth]{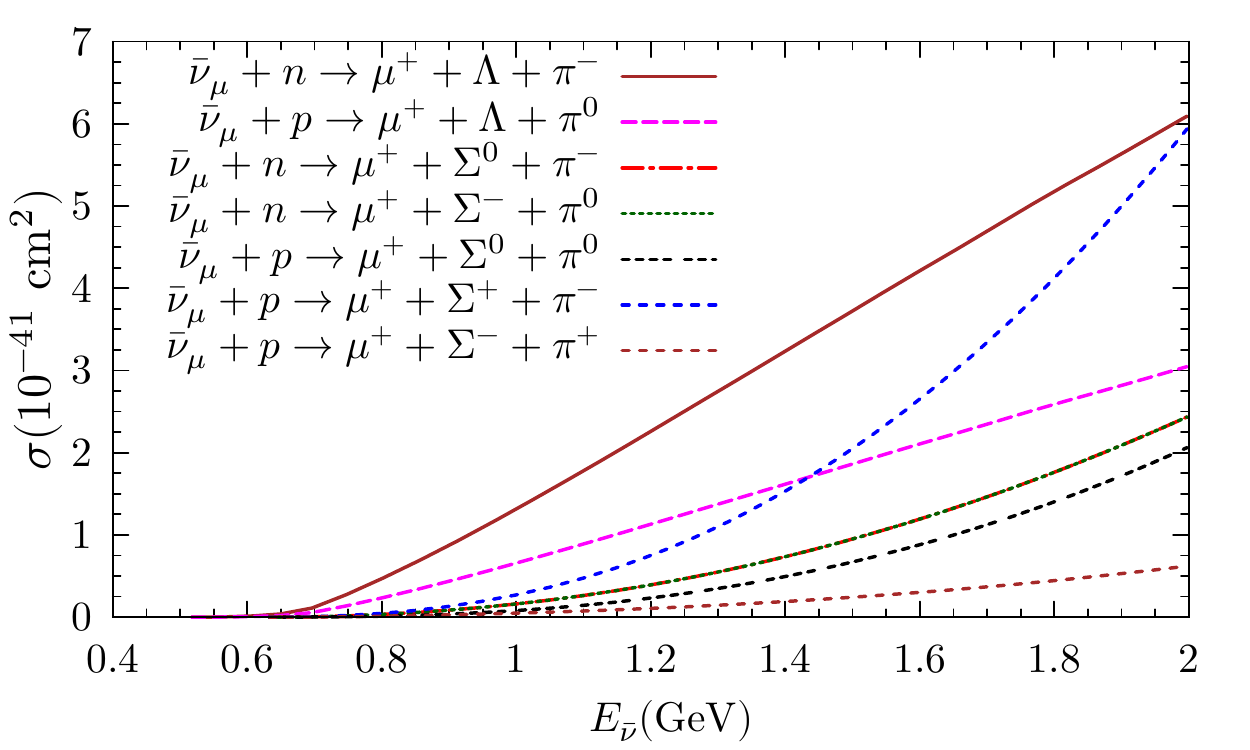}
  \caption{Plot of the total cross sections for $Y\pi$ production off
    nucleons induced by muon antineutrinos as a function of the
    antineutrino energy in the LAB frame.}
  \label{fig:numubar_total_xsect}
 \end{figure}
 
In Fig. \ref{fig:numubar_total_xsect}, we present the total cross
sections for the full model corresponding to all the possible $Y\pi$
channels induced by muon antineutrinos off nucleons as a function of
the antineutrino energy in the LAB frame.  It is interesting to see
that the the total cross sections have the same order of magnitude as
those of the single $K$ and $\bar{K}$ production ($1K/\bar K$) cross
sections off nucleons studied in Refs.
\cite{RafiAlam:2010kf,Alam:2012zz}.  While the $1K/\bar K$ cross
sections are smaller than the single pion cross sections because of
the smallness of the Cabibbo angle; the $Y\pi$ cross section misses
the strong $\Delta(1232)$-like mechanism, apart from the threshold
effect.
 
\begin{figure*}
\includegraphics[width=0.32\textwidth,height=.34 \textwidth]{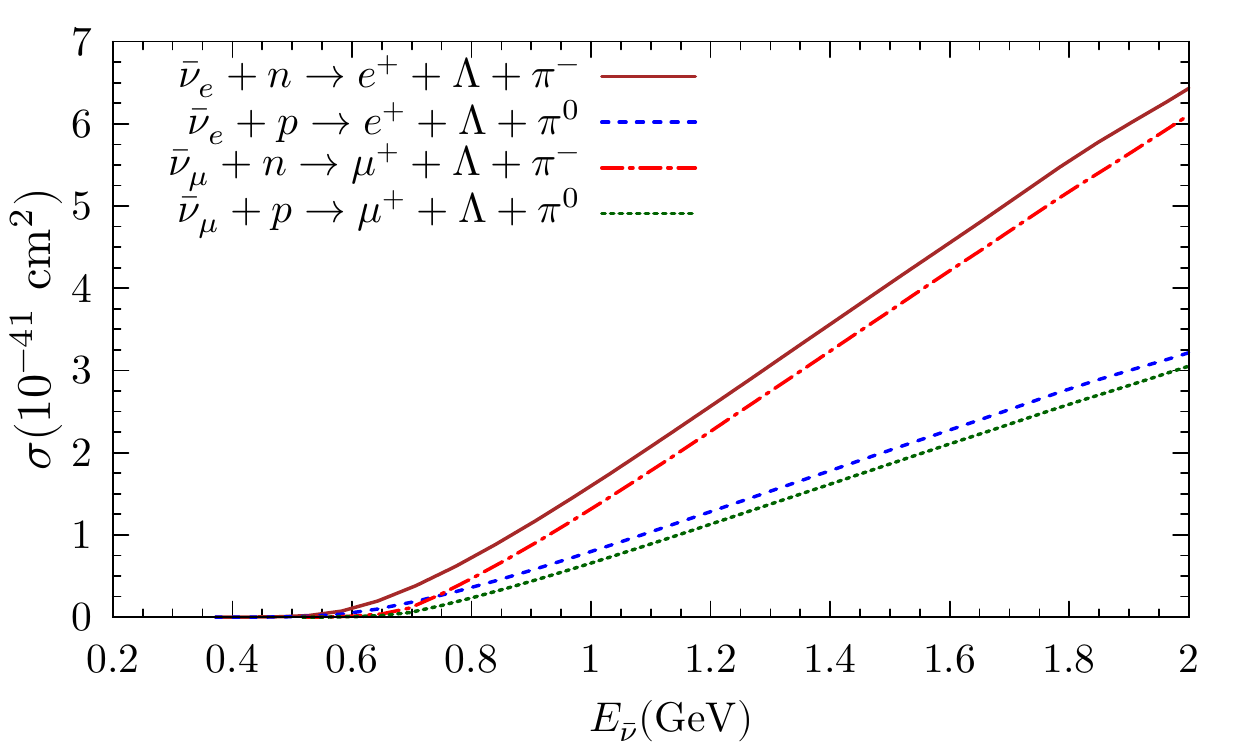}
\includegraphics[width=0.32\textwidth,height=.34 \textwidth]{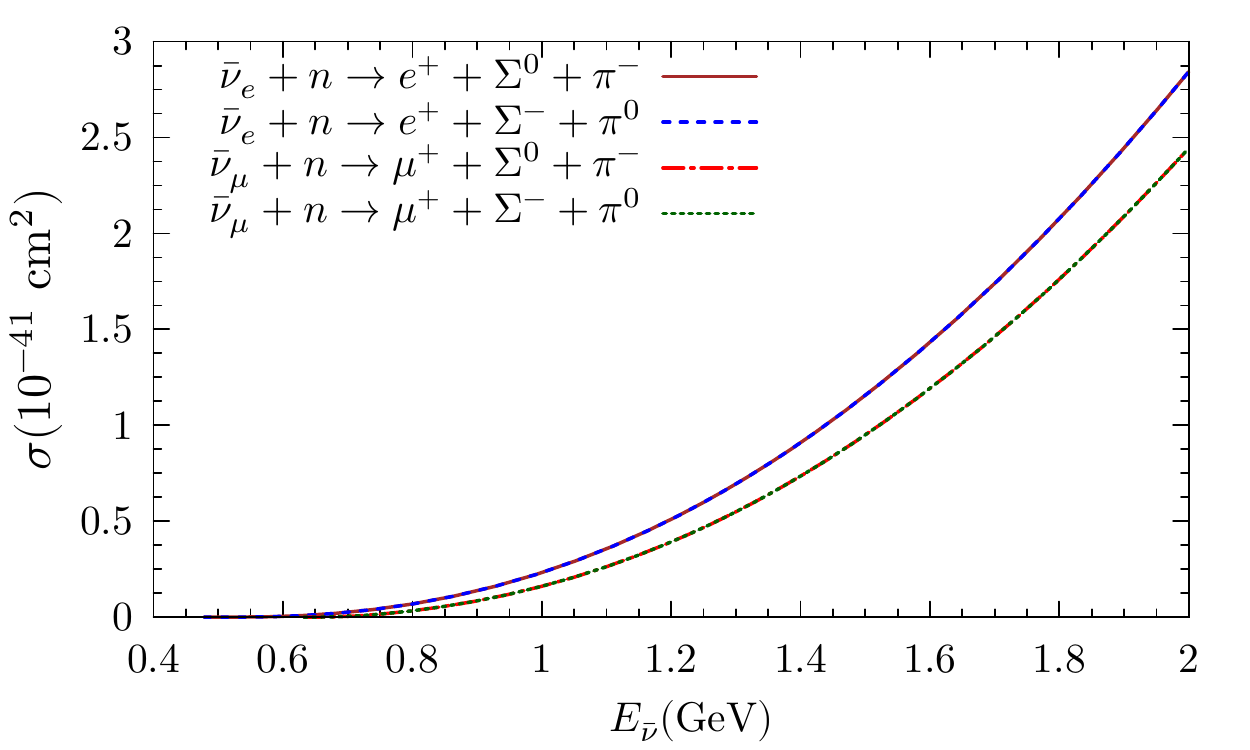}
 \includegraphics[width=0.32\textwidth,height=.34 \textwidth]{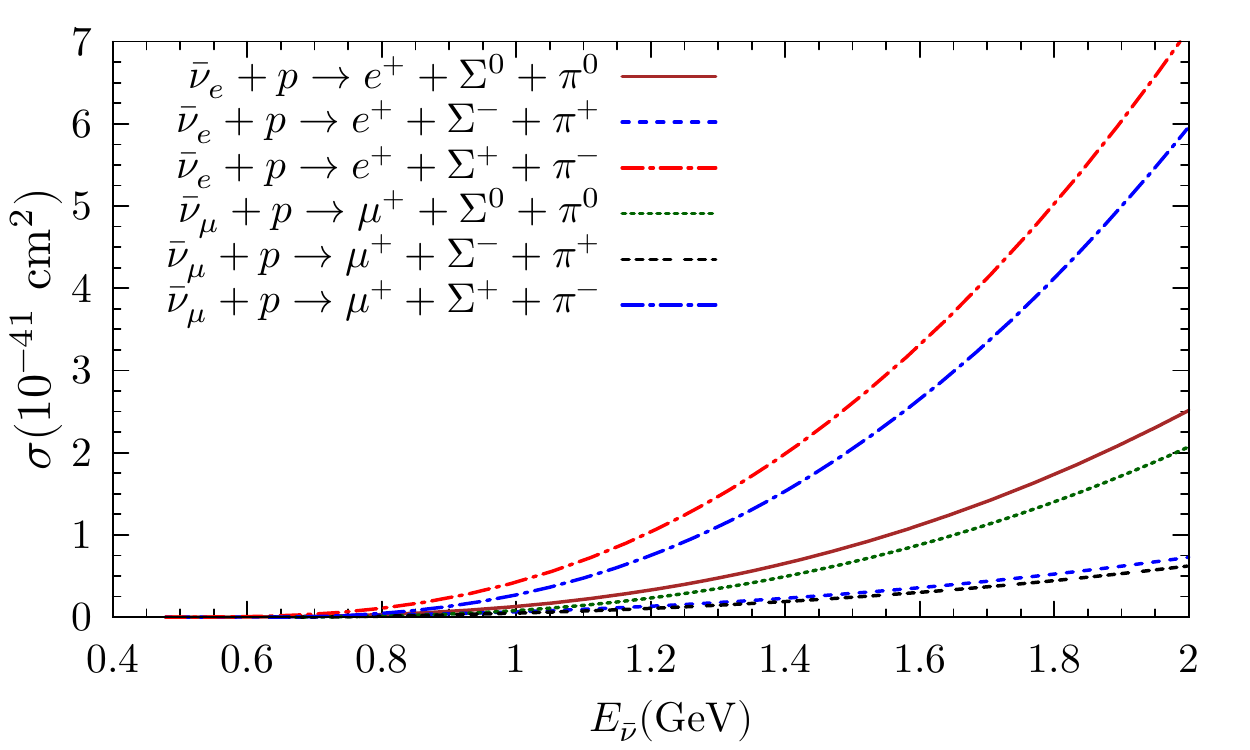}
\caption{Comparison between electron antineutrino and muon
  antineutrino induced total cross sections off nucleons in terms of
  the antineutrino energies in the LAB frame. In the left panel we
  display the $\Lambda\pi$ reaction channels.  In the middle panel we
  show the $\Sigma\pi$ production channels off neutrons. Finally, in
  the right panel we plot the $\Sigma\pi$ reactions off protons.}
\label{fig:electron_vs_muon_xsects}
\end{figure*}

Finally, in Fig. \ref{fig:electron_vs_muon_xsects} we show the
comparison between the electron antineutrino and muon antineutrino
induced $Y\pi$ production total cross sections as a function of the
antineutrino energy in the LAB frame.  As expected, the cross sections
for electron antineutrinos are larger than their muon counterparts
because of their lower production thresholds due to the smallness of
the final electron mass than the muon one.  A similar trend is found
for all other reaction channels.
 
 \subsection{Comparisons with other
 models}\label{subsec:comp_others}

\begin{figure*}
\begin{subfigure}[h]{0.5\textwidth}
\includegraphics[scale=0.65, bb = 5 0 340 240]{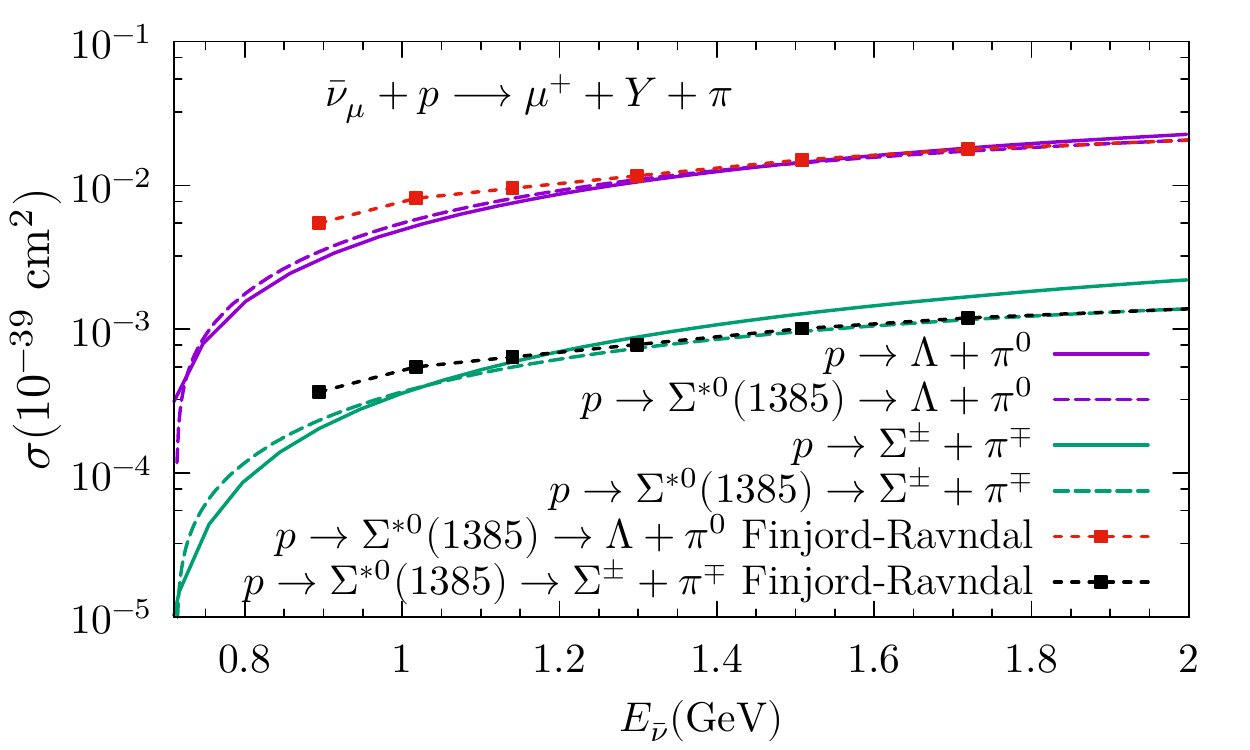}
\end{subfigure}
\begin{subfigure}[h]{0.49\textwidth}
\includegraphics[scale=0.65, bb = 5 0 340 240]{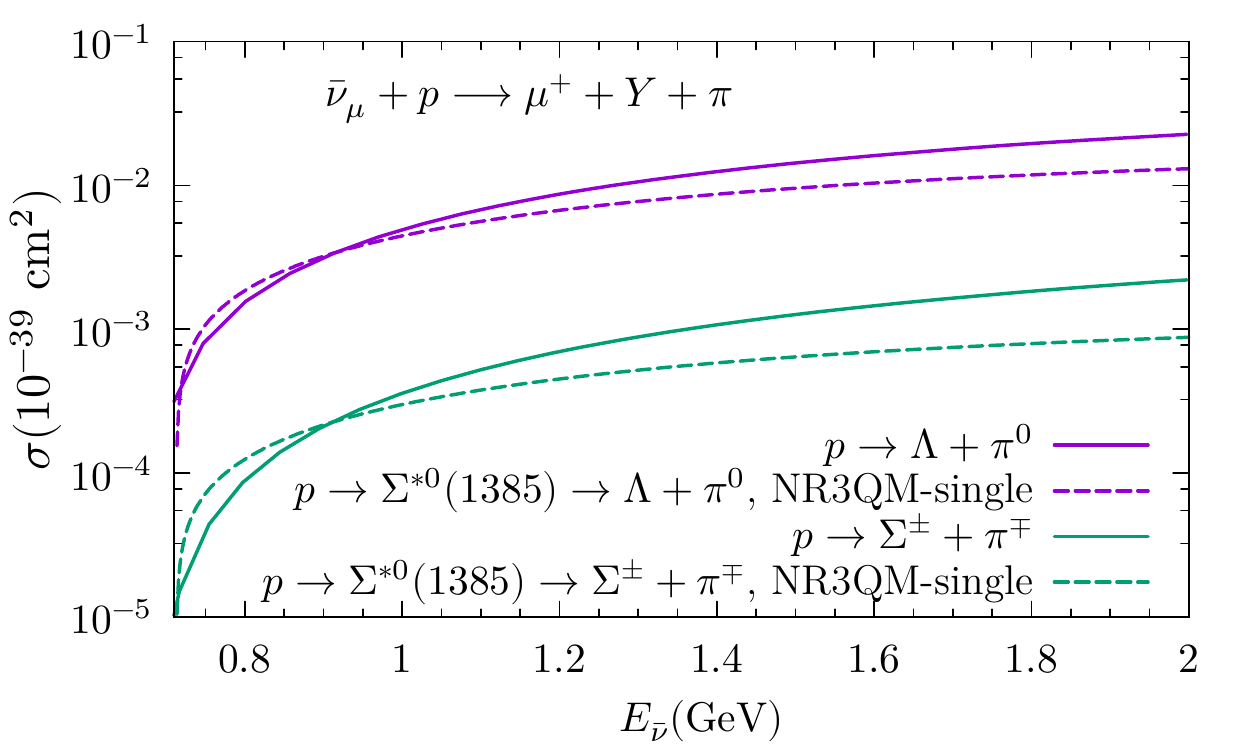}
\end{subfigure}
\caption{Comparison for the reaction of Cabibbo suppressed single pion
  production off protons with the mechanism of intermediate
  $\Sigma^{*0}$ alone. We compare with the results obtained in
  Ref. \cite{Wu:2013kla}, where the authors calculate the quasi-free
  production of an on-shell $\Sigma^{*0}$ off protons induced by muon
  antineutrinos. Solid lines represent our model with only
  s-$\Sigma^*$ reaction mechanism, while dashed lines are the results
  of Ref. \cite{Wu:2013kla} for the V-A approach(left panel); and the
  NR3QM-single approach(right-panel). On the left panel, we also
  display as dotted lines with filled squares the results of
  Ref. \cite{Finjord:1975zy}.}
\label{fig:numubar_xsect_few_body_syst}
\end{figure*}

This work presents a detailed analysis of the $Y\pi$ production cross
section induced by antineutrinos. To the best of our knowledge, our
calculations are one of the first in studying these processes.
However, there are independent calculations where the authors
calculate the quasi-free production of an on-shell $\Sigma^{*0}(1385)$
resonance~\cite{Wu:2013kla}.  In order to make a comparison with the
$\Sigma^{*0}(1385)$ production, we consider only the s-channel
$\Sigma^*$ diagram.  To compare the production cross sections of
specific $Y\pi$ channels, we have taken into account the primary decay
channels of $\Sigma^*$: $\Lambda\pi^0$ and $\Sigma\pi$ with branching
ratios 87\% and 11.7\% respectively~\cite{Zyla:2020zbs}.  Further, the
inclusive $\Sigma\pi$ decay channel may have different candidates,
viz, $\Sigma^{\pm}\pi^{\mp}$ and $\Sigma^0\pi^0$. The individual
contribution of these final states can be obtained by multiplying by
the appropriate (square of) Clebsch-Gordan coefficients, which is zero
for $\Sigma^0\pi^0$ and $\frac12$ for $\Sigma^{\pm}\pi^{\mp}$. The
results are shown in Fig. \ref{fig:numubar_xsect_few_body_syst}.  In
the left panel of Fig. \ref{fig:numubar_xsect_few_body_syst}, where
the two models show a remarkable coincidence, the solid lines
correspond to our model, while the dashed lines are those of
Ref. \cite{Wu:2013kla} with the V-A approach. They use an axial mass
$M_A=1.05$ GeV for the axial form factor $C^{A}_5(q^2)$, as being used
in the present model~\footnote{The readers should note that this axial
  mass used in the nucleon-to-resonance transition axial-vector form
  factor $C^A_5$ is different from the axial mass appearing in
  eq. (\ref{axial-vector-ff}) for the nucleon axial form
  factor.}. Off-shell effects present in our model show a slight
discrepancy at the highest energies shown in the left panel for the
$\Sigma^{\pm}\pi^{\mp}$ production channel. Also on this panel we show
as dotted lines with filled squares the corresponding results of
Ref. \cite{Finjord:1975zy}, where again the coincidence for the decay
channel $p\rightarrow\Sigma^{*0}(1385)\rightarrow\Lambda\pi^0$ at the
higher energies shown in the plot is remarkable.

In the right panel of Fig.  \ref{fig:numubar_xsect_few_body_syst}, we
compare our results with the non-relativistic 3 quark model
(NR3QM-single) discussed in Ref. \cite{Wu:2013kla}. In this case, the
discrepancies are larger at smaller antineutrino energies; however,
this is expected as the cross sections calculated within the
NR3QM-single approach were already smaller than those calculated
within the V-A approach (see Fig. 10 in Ref. \cite{Wu:2013kla}).
Nonetheless, we find that the cross sections are of the same order of
magnitude, even when comparing with the most unfavorable approach.

\begin{figure*}[htb]
\includegraphics[width=0.45\textwidth,height=.4 \textwidth]{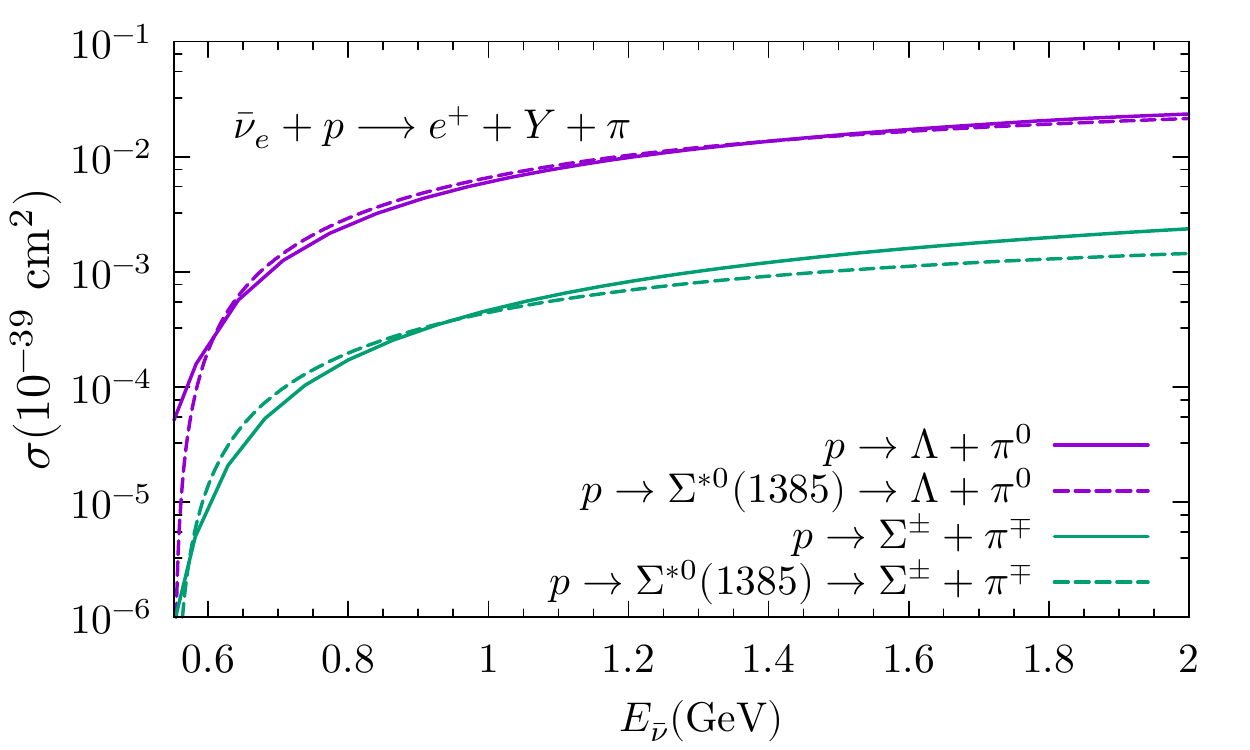}
\includegraphics[width=0.45\textwidth,height=.4 \textwidth]{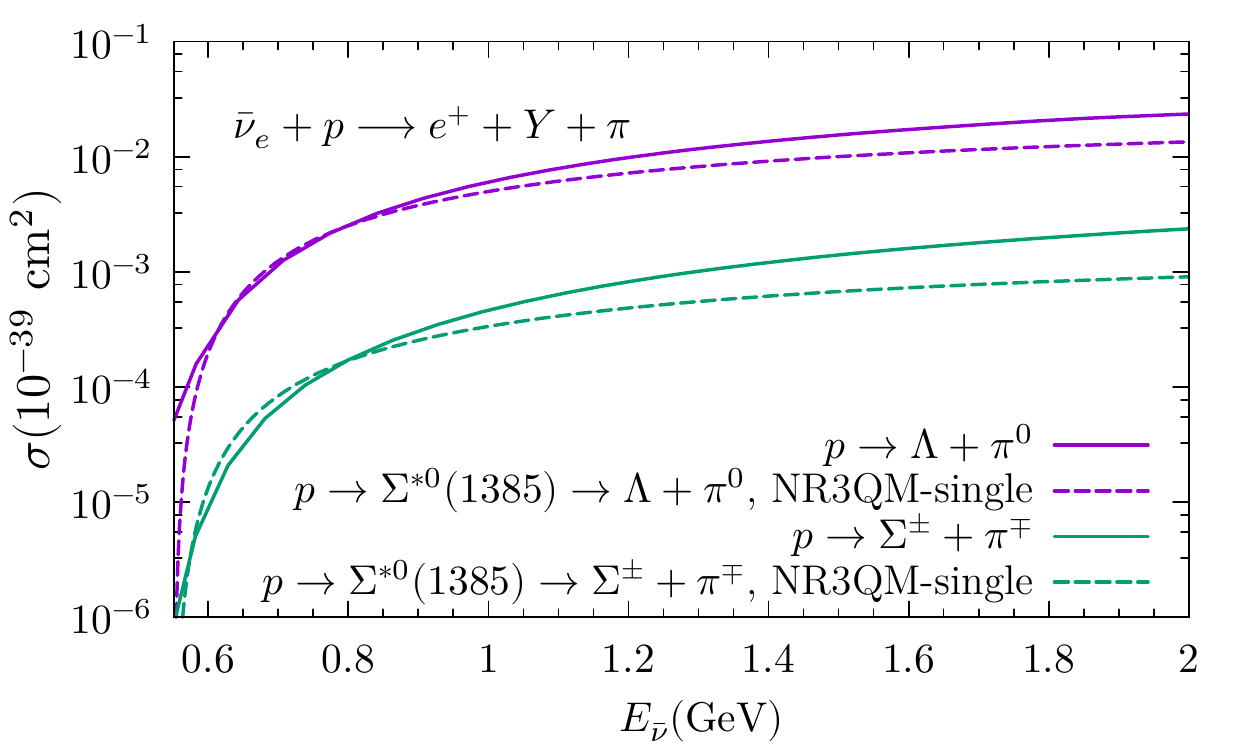}
\caption{Same as Fig. \ref{fig:numubar_xsect_few_body_syst} but for
  the reactions induced by electron antineutrinos off protons. Panels
  and lines have the same meaning as in Fig.
  \ref{fig:numubar_xsect_few_body_syst}.}
\label{fig:nuebar_xsect_few_body_syst}
\end{figure*}

Fig. \ref{fig:nuebar_xsect_few_body_syst} shows a similar comparison
as in Fig. \ref{fig:numubar_xsect_few_body_syst} but with the results
of Ref. \cite{Wu:2013kla} for the reactions induced by electron
antineutrinos off protons. In this latter case, the thresholds are a
bit lower, but the general features found in
Fig. \ref{fig:numubar_xsect_few_body_syst} remain the same.  One
should note that the comparison on the left panel of
Fig. \ref{fig:nuebar_xsect_few_body_syst} with the V-A approach of
Ref. \cite{Wu:2013kla} is expected as both models are identical,
except for the off-shell treatment of $\Sigma^{*0}(1385)$ resonance.
However, on the right panel of
Fig. \ref{fig:nuebar_xsect_few_body_syst}, the agreement with the
NR3QM-single approach is more inadequate as it already was in the
right panel of Fig. \ref{fig:numubar_xsect_few_body_syst}.

\begin{figure}[htb]
\includegraphics[width=0.45\textwidth,height=.4 \textwidth]{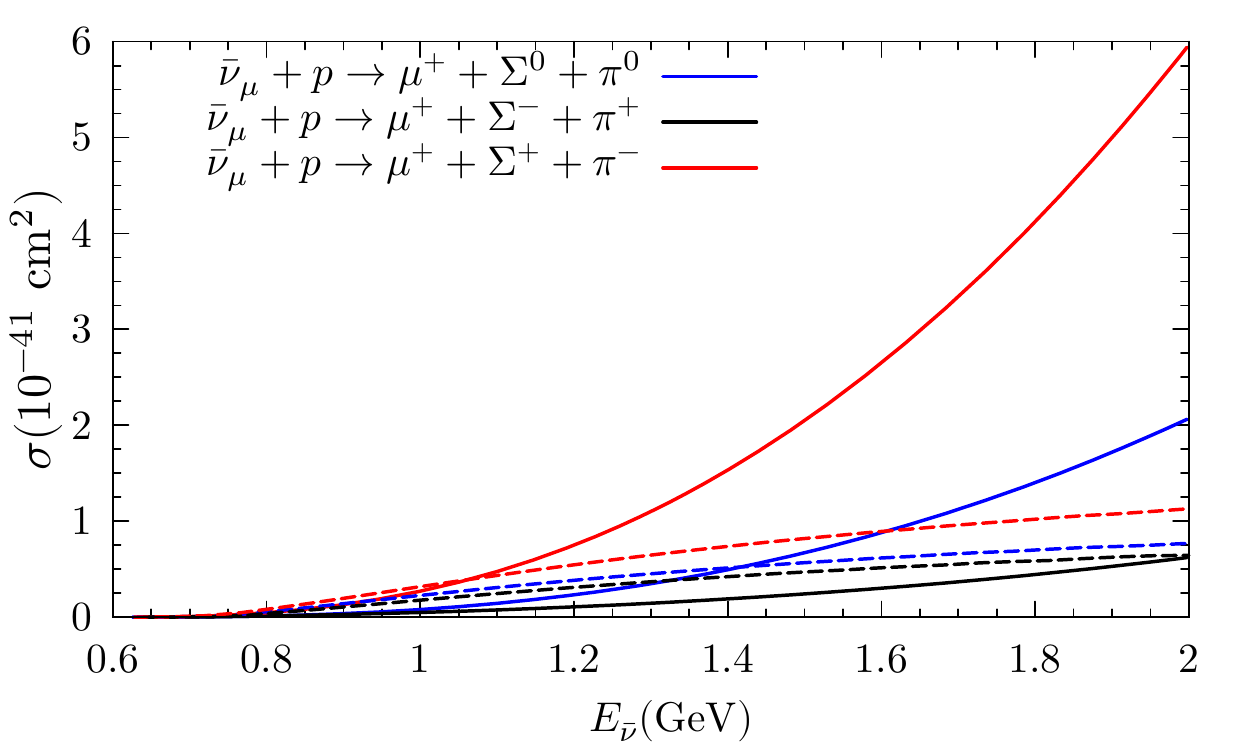}
\caption{Comparison between the total cross sections for the three
  $\Sigma\pi$ reaction channels for our model (solid lines) and that
  of Ref. \cite{Ren:2015bsa} (short-dashed lines).}
\label{fig:numubar_comparison_Luis}
\end{figure}

Finally, in Fig. \ref{fig:numubar_comparison_Luis} we show the
comparison between the results of the total cross sections for the
three charge $\Sigma\pi$ states production channels off protons in our
model (solid lines) versus the results of Ref. \cite{Ren:2015bsa}
(short-dashed lines). The model of Ref. \cite{Ren:2015bsa} is based on
a chiral unitary approach where all the meson-baryon pairs with $S=-1$
produced in a primary CT, KP or meson-in-flight (MF) diagram are
allowed to interact in a coupled channels approach to dynamically
generate the $\Lambda(1405)$ resonance by solving the Bethe-Salpeter
equation with an interaction potential derived from the lowest-order
chiral Lagrangian of eq. (\ref{eq:lag_mb}).

If we inspect Fig. \ref{fig:numubar_comparison_Luis}, we observe that
the total cross sections derived in our model are generally much
larger than those of Ref.  \cite{Ren:2015bsa}, especially important is
the enhancement for the $\Sigma^{+}\pi^{-}$ channel, which amounts to
almost a factor $6$ at $E_{\bar{\nu}}=2$ GeV. More moderate is the
enhancement for the $\Sigma^0\pi^0$ channel, and for the
$\Sigma^{-}\pi^{+}$ one, where our cross section is smaller than its
counterpart of Ref. \cite{Ren:2015bsa}.  Nonetheless, close to
threshold the three cross sections are larger in
Ref. \cite{Ren:2015bsa} than in our model, even although we explicitly
incorporate a resonant diagram with a $\Sigma^*(1385)$ resonance which
is below the $\Lambda(1405)$ resonance and above the $\Sigma\pi$
threshold. This clearly means that the $\Lambda(1405)$ plays an
important role in the description of these reactions close to
threshold for the $\Sigma\pi$ production channels. Probably the reason
for this is that the $\Lambda(1405)$ appears in s-wave coupled
channels and these are going to be much more important close to
threshold. However, the $\Sigma^*(1385)$ is a p-wave resonance like
the $\Delta$, and its contribution, already small due to its couplings
(as shown in Fig. \ref{fig:proton_sigma_xsect}) for these reactions,
starts to contribute more at higher antineutrino energies.

However, the $\Lambda(1405)$ resonance is not going to play any role
for the $\Lambda\pi^0$ production off protons because it appears in
the $I=0$ channel and the final one has $I=1$. In the coupled channel
approach of Ref. \cite{Ren:2015bsa} there is the possibility of
producing a final $\Lambda\pi^0$ through a loop of $\bar{K}N$
intermediate states produced in the weak vertices coupled to
$I=1$. And indeed, these $\bar{K}N$ states couple directly (at the
level of $V_{ij}$ in the nomenclature of Ref. \cite{Oset:1997it}) to
$\Lambda\pi^0$ (see $C_{ij}$ coefficients of table I of
\cite{Oset:1997it} for the couplings of the two $\bar{K}N$ states to
$\Lambda\pi^0$). Based on these arguments, we think the most reliable
and unaffected by the presence of higher lying strange resonances are
those reaction channels with a $\Lambda$-particle as a final state.

One similarity between the results of Ref. \cite{Ren:2015bsa} and ours
is that the order of the channels with larger cross sections matches
significantly, i.e., the cross section for $\Sigma^{+}\pi^{-}$
production is larger than that for $\Sigma^{0}\pi^{0}$ followed by
$\Sigma^{-}\pi^{+}$ production, and the above trend is consistent in
both approaches.  This extends the reliability in the present model.

Also note that in the calculations of Ref. \cite{Ren:2015bsa}, a
non-relativistic reduction of the amplitudes was carried out.  These
approximations can also have an impact in the differences observed in
the size of the cross sections for the same range of antineutrino
energies shown in Fig. \ref{fig:numubar_comparison_Luis}.  However, we
cannot at the present moment quantify how much of the difference comes
from the non-relativistic approximation and/or from other relevant
ingredients present in the model of Ref. \cite{Ren:2015bsa} and absent
in ours, or vice versa.

Finally, it is also worth noticing that the way these cross sections
rise in our model is very similar to how the crossed or u-channel
diagrams do it, especially the crossed $\Delta$ diagrams plotted in
Fig.  \ref{fig:proton_sigma_xsect}, which are very relevant by
themselves, especially for the $\Sigma^{+}\pi^{-}$ and
$\Sigma^{0}\pi^{0}$ reaction channels, which are those with the
largest cross sections. This could point to the importance of crossed
diagrams, not only for $\Delta$ intermediate states, but also for
$N^*$ resonances not considered here.

\subsection{Flux-integrated total cross sections}
\label{subsec:flux-folded total_xsects}

\begin{table*}[htb]
\begin{tabular}{|c|c|c|c|c|c|}
\hline
Reaction  & MiniBooNE & SciBooNE & T2K ND280 & T2K SK & Minerva\\
\hline
$\bar{\nu}_\mu + p \rightarrow \mu^{+} + \pi^0 + \Lambda$ & 3.42 & 
1.95 & 2.17 & 1.68 & 23.8 \\
\hline
$\bar{\nu}_\mu + n \rightarrow \mu^{+} + \pi^{-} + \Lambda$ & 6.84 & 3.90 & 4.33 
& 3.36 &47.7\\
\hline
$\bar{\nu}_\mu + p \rightarrow \mu^{+} + \pi^{0} + \Sigma^0$ & 0.935 & 0.713 & 
0.0684 & 0.0546 &0.623\\
\hline
$\bar{\nu}_\mu + p \rightarrow \mu^{+} + \pi^{-} + \Sigma^{+}$ & 2.88 & 2.13 & 
0.290& 0.231 &2.85\\
\hline
$\bar{\nu}_\mu + p \rightarrow \mu^{+} + \pi^{+} + \Sigma^{-}$ & 0.369 & 0.254 & 
0.111 & 0.0887 &1.36\\
\hline
$\bar{\nu}_\mu + n \rightarrow \mu^{+} + \pi^{-} + \Sigma^{0}$ & 1.38 & 0.954 & 
0.263 & 0.211 &2.96\\
\hline
$\bar{\nu}_\mu + n \rightarrow \mu^{+} + \pi^{0} + \Sigma^{-}$ & 1.38 & 0.954 &
0.263 & 0.211 &2.96\\
\hline
\end{tabular}
\caption{Flux-folded total cross sections for $\bar{\nu}_{\mu}$ fluxes
  from different experiments, in units of $10^{-42}$ cm${}^2$. The cut
  in the final invariant hadronic mass $W\leqslant1.4$ GeV has been
  applied to the calculations for the T2K and Minerva fluxes.  The
  uncertainties are in the last significant figure.}
\label{tab:flux-folded_xsects}
\end{table*}

\begin{figure*}[htb]
\includegraphics[width=0.45\textwidth,height=.4 \textwidth]{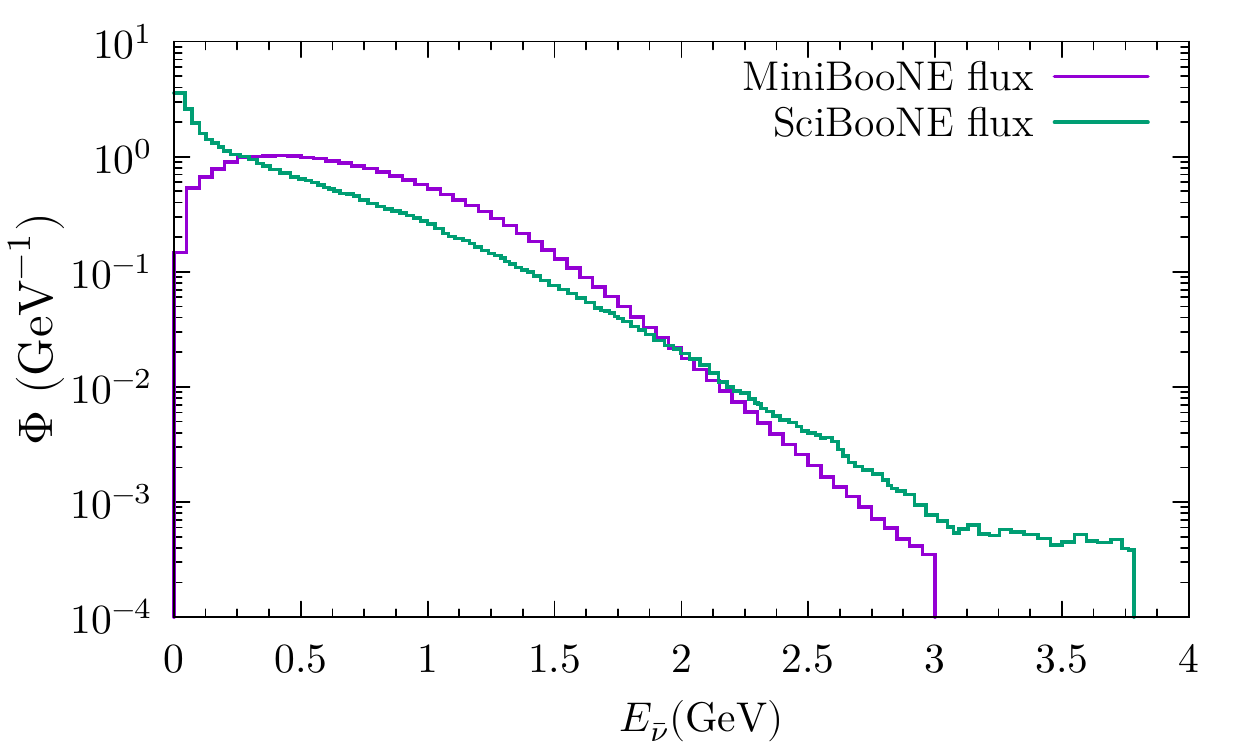}
\includegraphics[width=0.45\textwidth,height=.4 \textwidth]{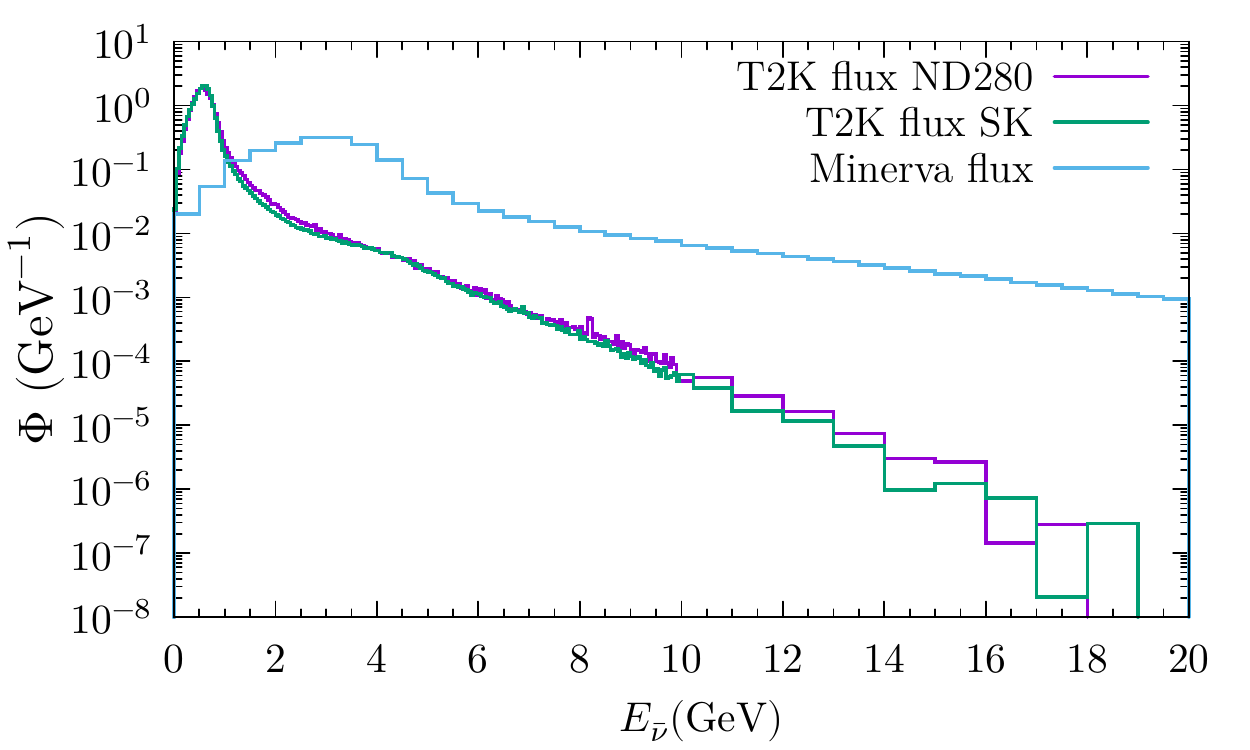}
\caption{Fluxes from different experiments.  On the left panel, the
  $\bar{\nu}_\mu$ fluxes from MiniBooNE \cite{Aguilar-Arevalo:2013dva}
  and SciBooNE \cite{Hiraide:2008eu}.  On the right panel, the T2K
  fluxes at the near detector ND280 and at the Super-Kamiokande far
  detector \cite{Abe:2012av,Abe:2015awa}, and the enriched
  $\bar{\nu}_\mu$ Minerva flux \cite{Aliaga:2016oaz}. The fluxes are
  normalized to their total flux, i.e, the integral of the fluxes
  shown in this figure is $1$.}
\label{fig:fluxes_experiments}
\end{figure*}

In this work, we have also estimated the flux-folded total cross
sections for antineutrino fluxes of several experiments like MiniBooNE
\cite{Aguilar-Arevalo:2013dva}, SciBooNE \cite{Hiraide:2008eu}, T2K
\cite{Abe:2012av,Abe:2015awa} and Minerva \cite{Aliaga:2016oaz}. The
energy dependence of these fluxes is shown in
Fig. \ref{fig:fluxes_experiments}.  We choose antineutrino fluxes that
peak at intermediate energies, i.e., $\left\langle E_{\bar{\nu}}
\right\rangle\simeq 1-3$ GeV.  At these energies, the four-momentum
transfers are expected to be low enough to carry chiral expansions,
making the present model more reliable.

The definition of the flux-integrated total cross section,
$\left\langle \sigma\right\rangle$, for a given antineutrino flux
$\Phi(E_{\bar{\nu}})$ of some experiment, can be obtained as
\begin{equation}\label{eq:sigma-flux-folded}
\left\langle \sigma \right\rangle=
\frac{\int^{E^{max}}_{E^{\rm th}_{\bar{\nu}}}\;\Phi(E_{\bar{\nu}})\;
\sigma(E_{\bar{\nu}})\;dE_{\bar{\nu}}}{\int^{E^{max}}_0\; \Phi(E_{\bar{\nu}})\;
dE_{\bar{\nu}}}.
\end{equation}
In eq. (\ref{eq:sigma-flux-folded}), the lower limit in the integral
of the numerator can be also zero, but it is not necessary, because
the total cross section $\sigma(E_{\bar{\nu}})$ is zero for
$E_{\bar{\nu}}<E^{\rm th}_{\bar{\nu}}$, where $E^{\rm th}_{\bar{\nu}}$
is the threshold antineutrino energy in the LAB frame for the reaction
to take place. Its expression is given by
\begin{equation}
E^{\rm th}_{\bar{\nu}}=\frac{\left(M_Y+m_\pi+m_l\right)^2 - M^2}{2M},
\end{equation}
thus giving $E^{\rm th}_{\bar{\nu}}\simeq0.515$ GeV for final
$\Lambda$ production and $E^{\rm th}_{\bar{\nu}}\simeq0.630$ GeV for
final $\Sigma$ production induced by muon antineutrinos.  While, the
$E^{max}$ depend upon the flux and their values are $20$ GeV for
Minerva and T2K, $3$ and $4$ GeV for MiniBooNE and SciBooNE,
respectively.

In table \ref{tab:flux-folded_xsects} we show the flux-folded total
cross sections for muon antineutrinos fluxes from different
experiments: MiniBooNE \cite{Aguilar-Arevalo:2013dva}, SciBooNE
\cite{Hiraide:2008eu}, T2K \cite{Abe:2012av,Abe:2015awa}, and Minerva
\cite{Aliaga:2016oaz}.

The T2K (both at the near detector ND280 and at Super-Kamiokande one)
and Minerva fluxes have larger tails ranging up to 20 GeV.  Our model,
which is based on a chiral expansion, is not going to be reliable for
these higher energies, where high momentum transfers and high
invariant masses become accessible with the increase of the
antineutrino energies.  In order to overcome this difficulty, we have
put a constraint on the final invariant hadronic mass, $W < 1.4$
GeV. This solves two problems: on the one hand, we are sure that
higher lying strange resonances above the $\Sigma^*(1385)$, such as
the $\Lambda(1405)$ (which has been shown in
Fig. \ref{fig:numubar_comparison_Luis} to contribute significantly to
the $\Sigma\pi$ production channel near threshold), are not going to
contribute for these kinematically constrained total cross sections;
on the other hand, the total cross sections when the cut in the
invariant mass is imposed, do not grow rapidly and hence allow to
calculate a well-defined flux-averaged total cross section with the
low energy fluxes like T2K and Minerva (low energy mode). In addition,
this cut has also a virtue, because it can be also experimentally
imposed, thus rejecting the $Y\pi$ events with measured invariant
masses $W>1.4$ GeV.

In order to analyze the results shown in table
\ref{tab:flux-folded_xsects}, it is important to remark that the
flux-folded total cross sections do not depend on the total flux,
because they are normalized to it. They depend basically on the shape
of the flux and where they are mostly peaked and if their tails are
longer or shorter. And also on how large is the total cross section in
the zone where the flux is sizeable. With this in mind, we can
understand the calculations shown in table
\ref{tab:flux-folded_xsects}.

The first comparison we analyze is between the flux-averaged cross
sections for MiniBooNE \cite{Aguilar-Arevalo:2013dva} and SciBooNE
\cite{Hiraide:2008eu} experiments.  Note that the flux taken for
MiniBooNE, Ref. \cite{Aguilar-Arevalo:2013dva}, corresponds to the
antineutrino enhanced sample, while the flux taken from figure 1 of
Ref. \cite{Hiraide:2008eu} corresponds also to the $\bar{\nu}_\mu$
flux, but in this case this is not the larger component of the flux,
because the latter is the muon neutrino component.

From table~\ref{tab:flux-folded_xsects}, we find that the results from
MiniBooNE and SciBooNE differ significantly, though the fluxes do not
look strikingly different in nature, see left panel of
Fig. \ref{fig:fluxes_experiments}. The reason for the differences is
that the SciBooNE flux peaks at antineutrino energies below the
threshold for the reaction to take place.  However, the SciBooNE flux
has a longer tail which decreases a bit slowly than the MiniBooNE
one. The flux averaged cross sections are always higher for MiniBooNE
than for SciBooNE because the MiniBooNE flux is larger in the region
between $0.5$ and $2$ GeV, and the presence of the SciBooNE tail has
little importance (specially for the $\Lambda\pi$ production channels)
even although in this region the cross section is growing (without the
cut in the hadronic invariant mass).

It is worth noting that there is a difference between the averaged
cross sections for the reactions $p\rightarrow\Lambda\pi^0$ and
$p\rightarrow\Sigma^{+}\pi^{-}$ in both experiments. The first
reaction has a higher flux-folded cross section with the MiniBooNE
flux, while the opposite happens with the SciBooNE one. The reason for
this has to be looked for in the behavior of the cross sections for
these two reactions in the higher energy tails of the fluxes. Indeed,
the $p\rightarrow\Sigma^{+}\pi^{-}$ cross section grows clearly
steeper with the antineutrino energy than the
$p\rightarrow\Lambda\pi^0$ one does, as can be seen in
Fig. \ref{fig:numubar_total_xsect}.  Therefore, the SciBooNE slowly
decreasing tail has a compensating effect for the
$p\rightarrow\Sigma^{+}\pi^{-}$ reaction, because in the region of
this tail, the cross section for $p\rightarrow\Sigma^{+}\pi^{-}$ is
much larger than that for the $p\rightarrow\Lambda\pi^0$ channel, thus
making the flux-folded $p\rightarrow\Sigma^{+}\pi^{-}$ cross section
the second in magnitude for the SciBooNE flux, while it was the third
in size with the MiniBooNE one.

\begin{figure*}[htb] 
\includegraphics[width=0.49\textwidth,height=.4\textwidth]{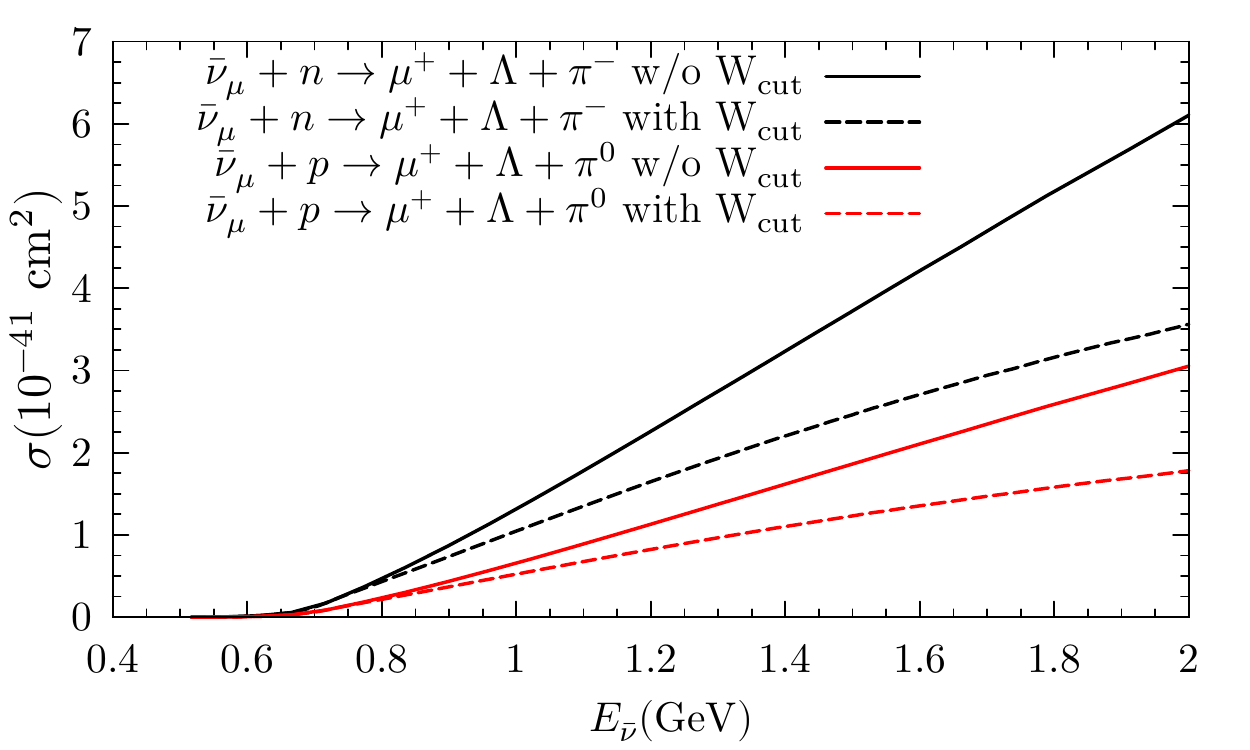} 
\includegraphics[width=0.49\textwidth,height=.4\textwidth]{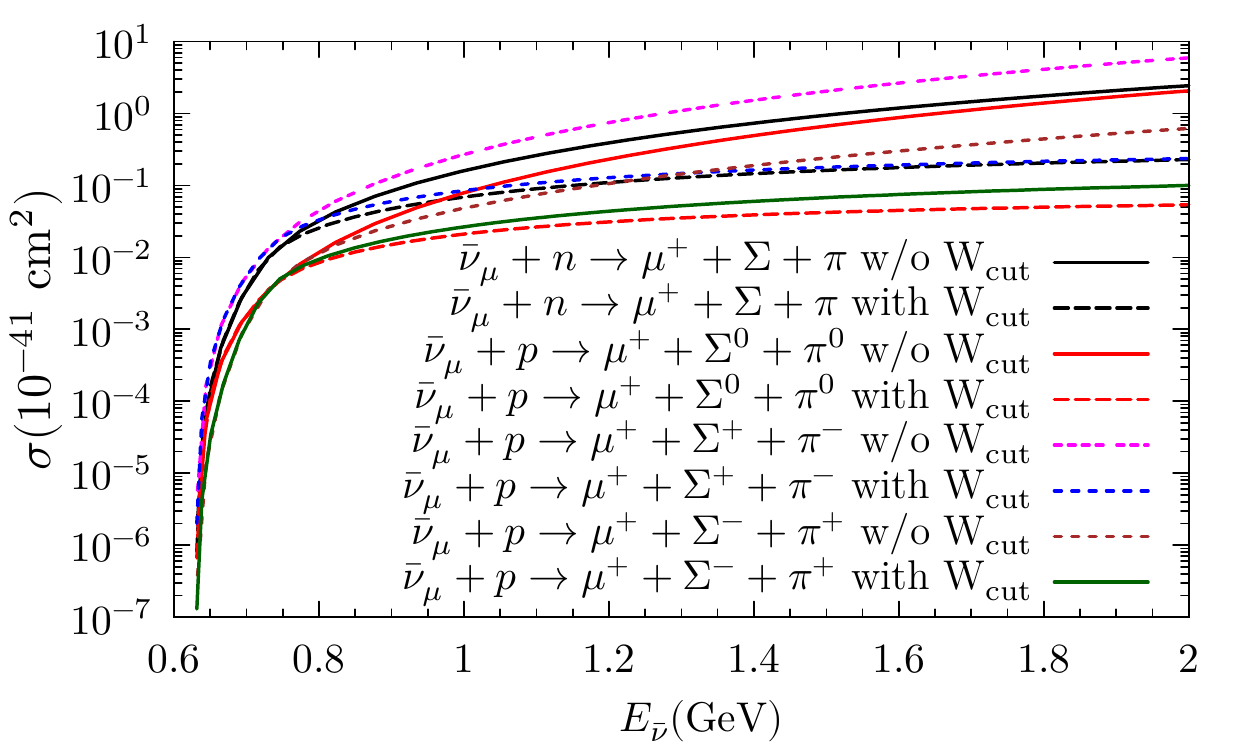} 
\caption{Plots of the total cross sections for the $Y\pi$ production
  as a function of the antineutrino energy with the effect of the
  kinematic cut in the final hadronic invariant mass $W\leq1.4$ GeV.
  In the left panel we show the results for $\Lambda\pi$ production,
  while in the right one we display those for the $\Sigma\pi$
  case. For this latter case the y-axis is logarithmic because of the
  huge reduction in the cross sections when the cut $W\le1.4$ GeV is
  imposed.}
\label{fig:xsects_with_Wcut}
\end{figure*}

For the flux-folded total cross sections with the T2K near detector
ND280, and Super-Kamiokande far detector fluxes
\cite{Abe:2012av,Abe:2015awa}, and with the Minerva flux
\cite{Aliaga:2016oaz}, we have applied the cut $W\le1.4$ GeV in the
final hadronic invariant mass.  This cut has the obvious effect of
reducing the size of the total cross sections, as can be observed in
Fig. \ref{fig:xsects_with_Wcut}.  However, the reduction in size is
much more prominent for the $\Sigma\pi$ reactions than for the
$\Lambda\pi$ ones. The reason for this behavior is because the cut in
the invariant mass is much closer to the threshold for $\Sigma\pi$
production ($W^{\Sigma\pi}_{\rm th}=M_\Sigma+m_\pi\simeq 1.33$ GeV)
than it is for the $\Lambda\pi$ production channels
($W^{\Lambda\pi}_{\rm th}=M_\Lambda+m_\pi\simeq 1.25$ GeV).  In fact,
obviously, if the applied cut had been below the $\Sigma\pi$
threshold, all these cross sections would have been exactly zero.

Therefore, this huge reduction in the size of the total cross sections
for the $\Sigma\pi$ reaction channels when the cut in the invariant
mass is applied explains why the flux-averaged total cross sections
with the T2K and Minerva fluxes are so small if compared with their
$\Lambda\pi$ counterparts in table \ref{tab:flux-folded_xsects}.  The
reduction due to the cut in the invariant mass amounts to a roughly
one order of magnitude smaller for the $\Sigma\pi$ reactions.  There
is even a reaction channel, $p\rightarrow\Sigma^0\pi^0$, where the
reduction of the cross section due to the cut in the invariant mass is
specially significant, as it can be observed in the right panel of
Fig. \ref{fig:xsects_with_Wcut}, because it is the smallest cross
section of the $\Sigma\pi$ channels, while this was not the case when
there was no cut in the final hadronic invariant masses. In fact, for
this particular reaction channel, the reduction in the flux-averaged
total cross sections is already two orders of magnitude than for the
$\Lambda\pi$ reactions. For this reason, we have plotted in
logarithmic scale the cross sections for the $\Sigma\pi$ channels when
comparing them with the cut and without it in the right panel of
Fig. \ref{fig:xsects_with_Wcut}, because in a linear vertical scale
the cross sections with the cut in the invariant mass were almost not
visible.

Of particular curiosity is the similarity of the flux-folded total
cross sections for the $p\rightarrow\Sigma^{+}\pi^{-}$ channel and the
$n\rightarrow\Sigma\pi$ (both final charge channels have exactly the
same cross section) one when the cut in the invariant mass is applied,
even for so different fluxes such as those of T2K and Minerva, which
are peaked at totally different antineutrino energies and have really
different tails, as shown in the right panel of
Fig. \ref{fig:fluxes_experiments}. However, as the reduced total cross
sections (due to the cut) for both channels are so similar (compare
blue and black dashed lines in the right panel of
Fig. \ref{fig:xsects_with_Wcut}), their flux-averaged total cross
sections shown in table \ref{tab:flux-folded_xsects} for the T2K and
Minerva fluxes are also very similar.  Nonetheless, the flux-averaged
cross section for the $p\rightarrow\Sigma^{+}\pi^{-}$ channel is
larger than those of the $n\rightarrow\Sigma\pi$ ones for the T2K
fluxes because these are peaked below $1$ GeV, where the cross section
for the $p\rightarrow\Sigma^{+}\pi^{-}$ production channel is a bit
larger.  For the Minerva flux the result is the opposite because this
flux is peaked around $3$ GeV, although the differences, as discussed,
are really minor.

It is also worth mentioning that even although both T2K fluxes at near
and far detectors are almost equal (see the right panel of Fig.
\ref{fig:fluxes_experiments}), the flux-folded total cross sections
are systematically smaller when convoluted with the flux at the SK
detector for all the reactions (the reader can compare the numbers in
the fourth and fifth columns of table
\ref{tab:flux-folded_xsects}). The reason for this has to be searched
in the slightly smaller tail of the T2K flux at SK, compared with that
at the ND280, especially in the region between $1$ and $4$ GeV of muon
antineutrino energies, where its contribution is still relevant for
the flux-integrated total cross section.

Finally, the large numbers for the flux-averaged total cross sections
with the Minerva flux shown in the last column of table
\ref{tab:flux-folded_xsects}, especially for the $\Lambda\pi$
production channels, and if compared with the same numbers for the T2K
fluxes, can be explained because the Minerva flux is peaked around $3$
GeV, where the cross sections are much larger than in the region where
the T2K fluxes are peaked. And, additionally, the larger and slowly
decreasing tail of the Minerva flux (solid cyan line on the right
panel of Fig.  \ref{fig:fluxes_experiments}) has also a very important
role in the enhancement of the flux-convoluted total cross sections
for this experiment, in comparison with the results obtained for T2K.

\section{Conclusions}\label{sec:conclusions}
In this work we have studied the Cabibbo suppressed single pion
production off nucleons induced by antineutrinos. This process, which
is the strangeness-changing counterpart of the largely studied single
pion production without change of strangeness, has been very scarcely
analyzed so far. In these reactions, the final pion is emitted along
with a $\Sigma$ or $\Lambda$ hyperon.

It is well-known that its Cabibbo enhanced counterpart is largely
driven by the weak excitation of the $\Delta$ resonance, therefore we
have also considered in our model the relevant ($S=-1$)
$\Sigma^*(1385)$ resonance, belonging to the same decuplet as the
$\Delta$. In fact, we have found that this mechanism is indeed the
dominant one for the $\Lambda\pi$ reactions, but of minor importance
for the $\Sigma\pi$ channels. We have also found that crossed $\Delta$
or nucleon-pole diagrams are also important, especially for some of
the $\Sigma\pi$ reactions. This could indicate that the inclusion of
$N^*$ resonances in the u-channel can be necessary, but the absence of
experimental data on these reactions refrains us from doing any
categorical statement about this.

We have also compared our results with others found in the recent and
past literature. The main conclusion is that the $\Lambda(1405)$
resonance plays an important role close to threshold, especially due
to its S-wave character, in comparison with the P-wave character of
the $\Sigma^*$ resonance. However, when one goes to higher
antineutrino energies, other mechanisms and higher partial waves start
to play an important role.  Because the $\Lambda(1405)$ is an isospin
$0$ state, we can say that this resonance is not going to have any
impact in the $\Lambda\pi$ reactions (which are those with the largest
total cross sections in our model up to antineutrino energies of $2$
GeV in the LAB frame) because there cannot be any coupling due to
conservation of strong isospin. Therefore, our most reliable results
are expected to be those producing final $\Lambda\pi$ hadrons for the
range of antineutrino energies explored in this work.

We have also studied the flux-convoluted total cross sections
of these reaction channels with the antineutrino fluxes of past
(MiniBooNE, SciBooNE) and current (T2K near and far detectors,
Minerva) neutrino oscillation and scattering experiments. The numbers
obtained for these flux-folded total cross sections, and given in
table \ref{tab:flux-folded_xsects}, together with the conclusions
drawn for the same observable with the antineutrino Minerva flux (also
with invariant mass cut) in table III of Ref. \cite{RafiAlam:2019rft},
indicate that these cross sections can be measured in Minerva
experiment, especially the cross sections for final $\Lambda\pi$
production.

Compared to $\Delta S =0$ pion production, the smallness of cross
section makes $\pi Y$ processes hard to detect. This means that the
feasibility of detecting these channels in experiments is also
limited. However, in some recent experiments like Minerva
\cite{Eberly:2014mra,Aliaga:2015wva,McGivern:2016bwh}, the
reconstructions of the incident neutrino/antineutrino energy and the
invariant hadronic mass were shown to be feasible for semi-inclusive
samples containing charged pions
\cite{Eberly:2014mra,McGivern:2016bwh} and a single neutral pion
\cite{Aliaga:2015wva} in charged current muon neutrino and
antineutrino scattering off hydrocarbon (CH) target, respectively.  In
fact, the experimental data for the total cross section as a function
of the antineutrino energy for the single neutral pion sample was
shown in Fig. 10b of reference \cite{McGivern:2016bwh}. In the lowest
energy bin, the cross section has a value of $19.8\times10^{-41}$
cm${}^2$/nucleon, although with more than 100\% of uncertainty.  For
larger energy bins, the uncertainties are much smaller.  Nevertheless,
we can also provide our results for larger antineutrino energies,
where the experimental data are expected to be less uncertain. The
caveat here is that we have to apply the cut in the invariant mass to
ensure that our model for the primary interaction is more reliable.
Moreover, the cross section is comparable at these energies for the
$\pi^0$ production channels; for example, see
Fig. \ref{fig:numubar_total_xsect}.  Our values shown in this figure
are about one order of magnitude smaller than that for the
semi-inclusive process studied in \cite{McGivern:2016bwh}. Thus, with
higher statistics we think they can be measured experimentally. Finally,
FSI experienced by pions in nuclear targets can indeed distort the
final signal, changing the identity of the final pion through
mechanisms like charge exchange; however, they may get compensated by
the secondary pions produced from hyperons.  Detailed analysis must be
required where our results may be used as the input for the effects
like FSI.

The primary pions produced in the reactions studied here have a
significant probability of being absorbed in the nucleus, but the
hyperons are long-lived particles ($\tau \sim 10^{-10}$ s, except for
the $\Sigma^0$) with small widths even in the nuclear medium
\cite{Singh:2006xp,Fatima:2018wsy} (this is particularly true for the
$\Lambda$) and exit the nucleus decaying weakly into secondary pions
and nucleons. If these nucleons are below the experimental detection
threshold (and therefore there is no way of reconstructing the
invariant mass of the decaying hyperon), the final signal for the
whole process could be indistinguishable from other mechanisms for
pion production. This would contribute to the distortion of the
tagging of the different processes leading to pion production in
antineutrino-nucleus scattering.

It is also worth mentioning that, recently, a revival of detectors
(basically high-pressure time projection chambers with adequate
admixtures of argon and hydrogen-enriched gases such as
methane~\cite{Hamacher-Baumann:2020ogq}) with high-quality momentum
resolution and using the technique of the transverse momentum
imbalance~\cite{Lu:2015hea} has emerged with the claim of being
exquisite for the measurements of neutrino/antineutrino-hydrogen cross
sections, and the discrimination of these reaction channels from other
background nuclides present in the target material. The good point of
these detection techniques is to eliminate nuclear effects at the
price of being able to detect only final charged particles. If
finally, this kind of detector prevails, then it will be possible to
study neutrino/antineutrino cross sections off free protons with high
accuracies, such as some of the proposed and studied in this work,
particularly those where all the final particles are charged.

Finally, we think that our model can be suitable to be implemented in
the Monte Carlo event generators as the primary interaction, which can
then be used as an input to simulate the propagation of the $\pi Y$
pair inside the nuclear medium after incorporating the relevant
nuclear effects.  Nonetheless, SU(3) breaking effects can also be
applied to see its plausible outcomes on cross sections within some
model based approaches like Ref. \cite{Faessler:2008ix}, already
applied in the $K\Xi$ production channel studied in
Ref. \cite{RafiAlam:2019rft}.

\section{Acknowledgements}\label{sec:acknowledgements}
The authors wish to thank Prof. Bing-Song Zou and Dr. Jia-Jun Wu for
providing their theoretical results of their study for comparison with
ours.

We also want to specially thank Dr. Luis Alvarez-Ruso for providing
the results of his work (together with his collaborators in that work)
for comparison with our results in this paper. And we also wish to
thank him for very fruitful discussions that have helped us to
understand their results and their underlying model, in order to try
to discuss as accurately as possible the implications of their results
in comparison with ours.

The authors also thank Dr. Justo Mart\'in-Albo for drawing our
attention to the new proposals for detection of pure
(anti)neutrino-hydrogen events with high pressure time projection
chambers.

To carry out some of the numerical calculations of this work the
authors have used the resources of the scientific computing cloud
PROTEUS of the Instituto Interuniversitario Carlos I de F\'isica
Te\'orica y Computacional of the University of Granada, and therefore
acknowledge this computational facility.

This work has been partially supported by Spanish Ministry of Science
under grants No. FIS2017-85053-C2-1-P and PID2020-114767GB-I00, and by
the Junta de Andalucia (grant No. FQM-225).  M.B.G. also acknowledges
support from Spanish Ministry of Science under grant PRE2018-083794.
M.R.A thanks UGC-BSR Startup Research Grant(F-NO. 30/2015/BSR),
Government of India for partial support.

\appendix

\section{SU(3) relations between the amplitudes}
\label{sec:appendix_a}
In this appendix we derive the relations between the amplitudes
(currents) for the seven reaction channels discussed in this work
using SU(3) group theoretical arguments.

First of all, we have to make the assignments between the physical
states and the mathematical (or SU(3)) ones for the meson and baryon
states of the octet before applying the Wigner-Eckart
theorem. Besides, we have to identify the irreducible tensor operator
belonging to the $\left\lbrace 8 \right\rbrace$ representation of
SU(3) group that drives the strangeness-changing weak transition.

The strangeness-changing weak charged current (without the $V_{us}$
Cabibbo-Kobayashi-Maskawa matrix element) carries ``magnetic" quantum
numbers of SU(3) $\left(I,I_3,Y \right)= \left(\frac12,-\frac12,-1
\right)$, i.e, those quantum numbers of the $K^-$ or the $\Xi^-$.
This current operator can be written at the quark level as
\begin{eqnarray}
j^\mu_{\Delta S=-1}&=&\overline{Q}\gamma^\mu(1-\gamma_5)(F_4-iF_5)Q
\nonumber\\
&=&-\sqrt{2}\;\overline{Q}\; 
K^{\mu\,\left\lbrace 8 \right\rbrace}_{\left(\frac12,-\frac12,-1 \right)}\;Q,
\end{eqnarray}
where $K^{\mu\,\left\lbrace 8
  \right\rbrace}_{\left(\frac12,-\frac12,-1\right)}=
-\frac{1}{\sqrt{2}} \gamma^\mu(1-\gamma_5)(F_4-iF_5)$, with
$F_i=\frac{\lambda_i}{2}$ (being $\lambda_i$ the Gell-Mann matrices).
But in general $K^{\mu\,\left\lbrace 8
  \right\rbrace}_{\left(\frac12,-\frac12,-1\right)}$ is an irreducible
tensor current operator belonging to the $\left\lbrace 8
\right\rbrace$ representation of the SU(3) group carrying the SU(3)
quantum numbers of this representation explicitly written in the
subindex. Therefore, to this operator we can apply the Wigner-Eckart
theorem of SU(3) \cite{deSwart:1963pdg}. Therefore, from here onwards
we will work with this operator by assuming that we do not have quarks
any longer and that the vector and axial-vector Dirac and Lorentz
structure can be more complex than simply $\gamma^\mu(1-\gamma_5)$,
which is the structure at the quark level only.

For simplicity in the notation, we will write the strangeness-changing
current operator simply as
\begin{equation}\label{eq:def_strangeness_changing_current}
j^\mu_{\rm sc} \equiv j^\mu_{\Delta S=-1}=
-\sqrt{2}\;K^{\mu\,\left\lbrace 8 \right\rbrace}_{\left(\frac12,-\frac12,-1 \right)},
\end{equation}
and we will calculate all the transition matrix elements driven by the
above current between initial nucleon states and final $\Sigma\pi$ and
$\Lambda\pi$ states. To this end, we have to fix the phases between
the physical states and the mathematical ones for which the SU(3)
Clebsch-Gordan coefficients have been calculated \cite{McNamee:1964xq,
  deSwart:1963pdg} in order to appropriately use the Wigner-Eckart
theorem. For the physical states we have in our study, this phase
fixing convention for mesons and baryons is:
\begin{eqnarray}
&&\left| p \right\rangle = \left| \left\lbrace 8 \right\rbrace; \frac12,\frac12,1
\right\rangle \qquad \left| n \right\rangle = \left| \left\lbrace 8 \right\rbrace; 
\frac12,-\frac12,1 \right\rangle \nonumber\\
&& \left| \Sigma^{+} \right\rangle = - \left| \left\lbrace 8 \right\rbrace;
 1,1,0 \right\rangle \qquad \left| \Sigma^{0} \right\rangle = 
 \left| \left\lbrace 8 \right\rbrace; 1,0,0 \right\rangle \nonumber\\
 && \left| \Sigma^{-} \right\rangle = 
 \left| \left\lbrace 8 \right\rbrace; 1,-1,0 \right\rangle \qquad
  \left| \Lambda \right\rangle = 
 \left| \left\lbrace 8 \right\rbrace; 0,0,0 \right\rangle \nonumber\\
 &&\left| \pi^{+} \right\rangle = - \left| \left\lbrace 8 \right\rbrace;
 1,1,0 \right\rangle \qquad \left| \pi^{0} \right\rangle = 
 \left| \left\lbrace 8 \right\rbrace; 1,0,0 \right\rangle \nonumber\\
 && \left| \pi^{-} \right\rangle = 
 \left| \left\lbrace 8 \right\rbrace; 1,-1,0 \right\rangle,
 \label{eq:sign_convention_states}
\end{eqnarray}
where the convention here is to label the mathematical states as
$\left| \left\lbrace \mathbf{N} \right\rbrace; I, I_3,
Y\right\rangle$.

The next step is to calculate the transition matrix elements
$\left\langle Y\pi \right| j^\mu_{\rm sc} \left| N \right\rangle$. To
this end, it is completely necessary to express the tensor product
$\left| Y \pi \right\rangle$ in the coupled basis by using the
Clebsch-Gordan coefficients that can be found in Ref.
\cite{McNamee:1964xq}, taking care of the signs found in some physical
states of eq. (\ref{eq:sign_convention_states}).  For completeness, we
provide below these expressions, although we know they are
straightforward.
\begin{widetext}
\begin{eqnarray}
\left| \Lambda\pi^0 \right\rangle &=& \sqrt{\frac{3}{10}}
\left| \left\lbrace 27 \right\rbrace; 1,0,0 \right\rangle - \frac12 
\left| \left\lbrace 10 \right\rbrace; 1,0,0 \right\rangle
- \frac12 \left| \left\lbrace \overline{10} \right\rbrace; 1,0,0 \right\rangle
+ \sqrt{\frac15} \left| \left\lbrace 8 \right\rbrace; 1,0,0 \right\rangle 
\label{eq:lambda_pi0_state} \\
\left| \Lambda\pi^- \right\rangle &=& \sqrt{\frac{3}{10}}
\left| \left\lbrace 27 \right\rbrace; 1,-1,0 \right\rangle - \frac12 
\left| \left\lbrace 10 \right\rbrace; 1,-1,0 \right\rangle
- \frac12 \left| \left\lbrace \overline{10} \right\rbrace; 1,-1,0 \right\rangle
\nonumber\\
&+& \sqrt{\frac15} \left| \left\lbrace 8 \right\rbrace; 1,-1,0 \right\rangle\\
\left| \Sigma^{+}\pi^{-} \right\rangle &=& - \sqrt{\frac16} 
\left| \left\lbrace 27 \right\rbrace; 2,0,0 \right\rangle - \sqrt{\frac{1}{12}}
\left| \left\lbrace 10 \right\rbrace; 1,0,0 \right\rangle
+ \sqrt{\frac{1}{12}}
\left| \left\lbrace \overline{10} \right\rbrace; 1,0,0 \right\rangle - 
\sqrt{\frac13}  \left| \left\lbrace 8^\prime \right\rbrace; 1,0,0 \right\rangle
\nonumber\\
&+& \sqrt{\frac{1}{120}}\left| \left\lbrace 27 \right\rbrace; 0,0,0 \right\rangle
+\sqrt{\frac15} \left| \left\lbrace 8 \right\rbrace; 0,0,0 \right\rangle
-\sqrt{\frac18} \left| \left\lbrace 1 \right\rbrace; 0,0,0 \right\rangle\\
\left| \Sigma^{0}\pi^{0} \right\rangle &=&  \sqrt{\frac23} 
\left| \left\lbrace 27 \right\rbrace; 2,0,0 \right\rangle  
+ \sqrt{\frac{1}{120}}\left| \left\lbrace 27 \right\rbrace; 0,0,0 
\right\rangle
+ \sqrt{\frac15} \left| \left\lbrace 8 \right\rbrace; 0,0,0 \right\rangle
-\sqrt{\frac18} \left| \left\lbrace 1 \right\rbrace; 0,0,0 \right\rangle\\
\left| \Sigma^{-}\pi^{+} \right\rangle &=& - \sqrt{\frac16} 
\left| \left\lbrace 27 \right\rbrace; 2,0,0 \right\rangle + \sqrt{\frac{1}{12}}
\left| \left\lbrace 10 \right\rbrace; 1,0,0 \right\rangle 
- \sqrt{\frac{1}{12}}
\left| \left\lbrace \overline{10} \right\rbrace; 1,0,0 \right\rangle + 
\sqrt{\frac13}  \left| \left\lbrace 8^\prime \right\rbrace; 1,0,0 \right\rangle
\nonumber\\
&+& \sqrt{\frac{1}{120}}\left| \left\lbrace 27 \right\rbrace; 0,0,0 \right\rangle
+\sqrt{\frac15} \left| \left\lbrace 8 \right\rbrace; 0,0,0 \right\rangle
-\sqrt{\frac18} \left| \left\lbrace 1 \right\rbrace; 0,0,0 \right\rangle\\
\left| \Sigma^{0}\pi^{-} \right\rangle &=&  \sqrt{\frac12} 
\left| \left\lbrace 27 \right\rbrace; 2,-1,0 \right\rangle + \sqrt{\frac{1}{12}}
\left| \left\lbrace 10 \right\rbrace; 1,-1,0 \right\rangle 
- \sqrt{\frac{1}{12}}
\left| \left\lbrace \overline{10} \right\rbrace; 1,-1,0 \right\rangle \nonumber\\
&+& \sqrt{\frac13}  \left| \left\lbrace 8^\prime \right\rbrace; 1,-1,0 \right\rangle\\
\left| \Sigma^{-}\pi^{0} \right\rangle &=&  \sqrt{\frac12} 
\left| \left\lbrace 27 \right\rbrace; 2,-1,0 \right\rangle - \sqrt{\frac{1}{12}}
\left| \left\lbrace 10 \right\rbrace; 1,-1,0 \right\rangle
+ \sqrt{\frac{1}{12}}
\left| \left\lbrace \overline{10} \right\rbrace; 1,-1,0 \right\rangle \nonumber\\
&-&\sqrt{\frac13}  \left| \left\lbrace 8^\prime \right\rbrace; 1,-1,0 \right\rangle.
\label{eq:sigmaminus_pi0_state}
\end{eqnarray}
\end{widetext}

Now we calculate the matrix elements $\left\langle Y\pi \right|
j^\mu_{\rm sc} \left| N \right\rangle$ but expressing the bras
$\left\langle Y\pi \right|$ in terms of the coupled basis as given in
eqs. (\ref{eq:lambda_pi0_state})-(\ref{eq:sigmaminus_pi0_state}), and
then apply the Wigner-Eckart theorem to each matrix element because
now we have an irreducible tensor operator between states belonging to
irreducible representations of the SU(3) group. For completeness,
below we write the expression of the Wigner-Eckart theorem for SU(3),
which can be also found in \cite{deSwart:1963pdg},
\begin{eqnarray}
&&\left\langle \left\lbrace\mu_3\right\rbrace; (\nu_3) \right| 
T^{\left\lbrace \mu_2 \right\rbrace}_{(\nu_2)} \left| 
\left\lbrace\mu_1\right\rbrace; (\nu_1) \right\rangle = \nonumber\\
&&\sum_{\gamma} 
\left(
\begin{array}{ccc}
\left\lbrace \mu_1 \right\rbrace & \left\lbrace \mu_2 \right\rbrace & 
\left\lbrace \mu_3 \right\rbrace_{\gamma} \\
(\nu_1) & (\nu_2) & (\nu_3) \\
\end{array}
\right)
\left\langle \left\lbrace\mu_3\right\rbrace \right| | 
T^{\left\lbrace \mu_2 \right\rbrace} | \left| \left\lbrace\mu_1\right\rbrace
 \right\rangle_{\gamma}.\nonumber\\ \label{eq:WE_theorem}
\end{eqnarray}

In the above expression, the indices $\mu_i$ refer to the irreducible
representations of the SU(3) group, while the indices $\nu_i$
collectively refer to the $(I,I_3,Y)$ ``magnetic" quantum numbers of
the representation $\mu_i$. The factor between parentheses is
precisely the SU(3) Clebsch-Gordan coefficient, and finally the last
term in eq. (\ref{eq:WE_theorem}) is the reduced matrix element, which
is totally independent of the ``magnetic" quantum numbers. Note that,
in principle, a sum over $\gamma$ has to be carried out. This amounts
to sum over all the times the $\left\lbrace\mu_3\right\rbrace$
irreducible representation is contained in the tensor product
$\left\lbrace\mu_1\right\rbrace \otimes
\left\lbrace\mu_2\right\rbrace$.  However, in our case there will not
be such a sum because in the bras of eq. (\ref{eq:WE_theorem}) there
will always be a definite $\left\lbrace\mu_3\right\rbrace_{\gamma}$
representation.

After having evaluated the $\left\langle Y\pi \right| j^\mu_{\rm sc}
\left| N \right\rangle$ matrix elements for all the cases in our
study, we can write the following $7\times6$ matrix relating the
previous matrix elements with the reduced ones,
\begin{widetext}
\begin{equation}\label{eq:matrix_matrix_elements}
\left(
\begin{array}{c}
j^\mu_{p\rightarrow\Lambda\pi^0} \\
j^\mu_{n\rightarrow\Lambda\pi^{-}} \\
j^\mu_{p\rightarrow\Sigma^{+}\pi^{-}} \\
j^\mu_{p\rightarrow\Sigma^{0}\pi^{0}} \\
j^\mu_{p\rightarrow\Sigma^{-}\pi^{+}}  \\
j^\mu_{n\rightarrow\Sigma^{0}\pi^{-}}  \\
j^\mu_{n\rightarrow\Sigma^{-}\pi^{0}} 
\end{array}
\right)=
\left(
\begin{array}{cccccc}
\frac{\sqrt{3}}{10} & \frac{1}{\sqrt{48}} & \frac{-1}{\sqrt{48}} & 
\frac{-\sqrt{3}}{10} & 0 & 0 \\
\frac{\sqrt{3}}{\sqrt{50}} & \frac{1}{\sqrt{24}} & \frac{-1}{\sqrt{24}} & 
\frac{-\sqrt{3}}{\sqrt{50}} & 0 & 0 \\
\frac{1}{40} & \frac{1}{12} & \frac{1}{12} & \frac{1}{10} & \frac{-1}{6} & 
\frac{-1}{8} \\
\frac{1}{40} & 0 & 0 & \frac{1}{10} & 0 & \frac{-1}{8} \\
\frac{1}{40} & \frac{-1}{12} & \frac{-1}{12} & \frac{1}{10} & \frac16 & 
\frac{-1}{8} \\
0 & \frac{-1}{\sqrt{72}} & \frac{-1}{\sqrt{72}} & 0 & \frac{1}{\sqrt{18}} & 0 \\
0 & \frac{1}{\sqrt{72}} & \frac{1}{\sqrt{72}} & 0 &  \frac{-1}{\sqrt{18}} & 0
\end{array}
\right)
\left(
\begin{array}{c}
j^\mu_{\left\lbrace 27 \right\rbrace} \\
j^\mu_{\left\lbrace 10 \right\rbrace} \\
j^\mu_{\left\lbrace \overline{10} \right\rbrace} \\
j^\mu_{\left\lbrace 8 \right\rbrace} \\
j^\mu_{\left\lbrace 8^\prime \right\rbrace} \\
j^\mu_{\left\lbrace 1 \right\rbrace}
\end{array}
\right),
\end{equation}
\end{widetext}
where $j^\mu_{N\rightarrow Y\pi}$ is a shorthand notation for
$\left\langle Y\pi \right| j^\mu_{\rm sc} \left| N \right\rangle$,
while $j^\mu_{\left\lbrace \mathbf{N} \right\rbrace}$ is also a
shorthand notation for the reduced matrix element $\left\langle
\left\lbrace N \right\rbrace \right|\left| j^\mu_{\rm sc}
\right|\left| \left\lbrace 8 \right\rbrace \right\rangle$, with
$j^\mu_{\rm sc}$ given by
eq. (\ref{eq:def_strangeness_changing_current}) and $\left\lbrace N
\right\rbrace$ is any of the irreducible representations of the SU(3)
group appearing in the Clebsch-Gordan series of the tensor product of
two octets, given in eq. (\ref{eq:CG_series}).

Of course, the coefficient matrix of
eq. (\ref{eq:matrix_matrix_elements}) has more rows than columns,
because for these $\Delta S=-1$ weak strangeness-changing transitions
there are only $6$ independent matrix elements, $j^\mu_{\left\lbrace
  \mathbf{N} \right\rbrace}$. However, not 6 matrix elements of the
left-hand side of eq. (\ref{eq:matrix_matrix_elements}) can be chosen
as truly independent, because the rank of the coefficient matrix is
not $6$, it is lesser. This was expectable, because there are other
independent transition matrix elements that can be driven by the weak
strangeness-changing operator of eq.
(\ref{eq:def_strangeness_changing_current}). These can be, for
instance, the $\left\langle N^\prime \bar{K} \right| j^\mu_{\rm sc}
\left| N \right\rangle$ (studied in Ref. \cite{Alam:2012zz}), the
$\left\langle \Xi K \right| j^\mu_{\rm sc} \left| N \right\rangle$
(studied in Ref. \cite{RafiAlam:2019rft}), or the $\left\langle Y\eta
\right| j^\mu_{\rm sc} \left| N \right\rangle$ matrix elements.

Indeed, the rank of the coefficient matrix of eq.
(\ref{eq:matrix_matrix_elements}) is $3$. It is easy to realize that
the first and second rows of this matrix are proportional. If one
multiplies the second row by a factor $\frac{1}{\sqrt{2}}$, one
obtains the coefficients of the first row. This indicates that only
one of the $j^\mu_{p\rightarrow\Lambda\pi^0}$ or
$j^\mu_{n\rightarrow\Lambda\pi^{-}}$ can be taken as independent.  The
relation between them is
\begin{equation}\label{eq: amp_lamb_pi0_vs_lamb_pim}
\left\langle \Lambda\pi^0 \right| j^\mu_{\rm sc} \left| p \right\rangle=
\frac{1}{\sqrt{2}}
\left\langle \Lambda\pi^{-} \right| j^\mu_{\rm sc} \left| n \right\rangle.
\end{equation}
Due to this relation between the amplitudes for $\Lambda\pi$
production, the cross sections for $n\rightarrow\Lambda\pi^{-}$
channel are twice as large than those for the
$p\rightarrow\Lambda\pi^0$ one, as can be observed in
Fig. \ref{fig:lambda_xsect}.

Another easy to notice relation can be drawn by observing the last two
rows of the matrix of eq. (\ref{eq:matrix_matrix_elements}). One is
the negative of the other, thus implying that
\begin{equation}\label{eq: amp_sig0_pim_vs_sigm_pi0}
\left\langle \Sigma^0\pi^{-} \right| j^\mu_{\rm sc} \left| n \right\rangle=
-\left\langle \Sigma^{-}\pi^{0} \right| j^\mu_{\rm sc} \left| n \right\rangle.
\end{equation}
This is the reason because of the cross sections for $\Sigma\pi$
production reactions off neutrons are exactly the same, as discussed
in the caption of Fig. \ref{fig:neutron_sigma_xsect}, and also the
flux-averaged cross sections shown in the last two rows of table
\ref{tab:flux-folded_xsects}.

Nonetheless, we have decided to take as independent
strangeness-changing matrix elements $\left\langle \Lambda\pi^{-}
\right| j^\mu_{\rm sc} \left| n \right\rangle$, $\left\langle
\Sigma^{+}\pi^{-} \right| j^\mu_{\rm sc} \left| p \right\rangle$ and
$\left\langle \Sigma^{-}\pi^{+} \right| j^\mu_{\rm sc} \left| p
\right\rangle$.  This can be done because by taking the second, third
and fifth rows of the matrix in eq. (\ref{eq:matrix_matrix_elements}),
one can form a $3\times6$ sub-matrix with at least one $3\times3$
determinant different from zero, i.e, these rows are linearly
independent~\footnote{One could have taken equally other $3$ different
amplitudes with the same properties of linear independence, but we
have decided to make this choice.}. With this choice, we can express
three $j^\mu_{\left\lbrace \mathbf{N} \right\rbrace}$ reduced matrix
elements in terms of the above linearly independent explicit
amplitudes and the other three remaining reduced matrix
elements~\footnote{One cannot express the six $j^\mu_{\left\lbrace
  \mathbf{N} \right\rbrace}$ reduced matrix elements in terms only of
the three explicit linear independent amplitudes, because there are
more unknowns than linearly independent equations in the system, i.e,
it is an underdetermined linear system.}. The result is,
\begin{widetext}
\begin{eqnarray}
j^\mu_{\left\lbrace 8 \right\rbrace}&=&\frac56\left( 
j^\mu_{\left\lbrace 10 \right\rbrace}-
j^\mu_{\left\lbrace \overline{10} \right\rbrace}\right)+
j^\mu_{\left\lbrace 27 \right\rbrace}-5\sqrt{\frac23}\,
j^\mu_{n\rightarrow\Lambda\pi^{-}} \label{eq:jmu8}\\
j^\mu_{\left\lbrace 8^\prime \right\rbrace}&=&\frac12\left(
j^\mu_{\left\lbrace 10 \right\rbrace}+
j^\mu_{\left\lbrace \overline{10} \right\rbrace}\right)+3\left(
j^\mu_{p\rightarrow\Sigma^{-}\pi^{+}}-
j^\mu_{p\rightarrow\Sigma^{+}\pi^{-}}\right)\\
j^\mu_{\left\lbrace 1 \right\rbrace}&=&\frac23 \left( 
j^\mu_{\left\lbrace 10 \right\rbrace}-
j^\mu_{\left\lbrace \overline{10} \right\rbrace}\right)+
j^\mu_{\left\lbrace 27 \right\rbrace}-4\sqrt{\frac23}\,
j^\mu_{n\rightarrow\Lambda\pi^{-}}-4\left( 
j^\mu_{p\rightarrow\Sigma^{-}\pi^{+}}+
j^\mu_{p\rightarrow\Sigma^{+}\pi^{-}}\right).\label{eq:jmu1}
\end{eqnarray}
\end{widetext}

Finally, if we replace the expressions for $j^\mu_{\left\lbrace
  \mathbf{N} \right\rbrace}$ given in eqs.
(\ref{eq:jmu8})-(\ref{eq:jmu1}) in the right-hand side of the linear
system of eq. (\ref{eq:matrix_matrix_elements}), and carry out the
matrix multiplication, we obtain eq.  (\ref{eq:
  amp_lamb_pi0_vs_lamb_pim}) for the first row. And also
\begin{widetext}
\begin{eqnarray}
\left\langle \Sigma^0\pi^0 \right| j^\mu_{\rm sc} \left| p \right\rangle&=&
\frac12 \left( 
\left\langle \Sigma^{+}\pi^{-} \right| j^\mu_{\rm sc} \left| p \right\rangle+
\left\langle \Sigma^{-}\pi^{+} \right| j^\mu_{\rm sc} \left| p \right\rangle
\right)\\
\left\langle \Sigma^0\pi^{-} \right| j^\mu_{\rm sc} \left| n \right\rangle&=&
\frac{1}{\sqrt{2}}\left( 
\left\langle \Sigma^{-}\pi^{+} \right| j^\mu_{\rm sc} \left| p \right\rangle-
\left\langle \Sigma^{+}\pi^{-} \right| j^\mu_{\rm sc} \left| p \right\rangle
\right)\label{eq:n_sigma0_piminus}\\
\left\langle \Sigma^{-}\pi^{0} \right| j^\mu_{\rm sc} \left| n \right\rangle&=&
-\frac{1}{\sqrt{2}}\left( 
\left\langle \Sigma^{-}\pi^{+} \right| j^\mu_{\rm sc} \left| p \right\rangle-
\left\langle \Sigma^{+}\pi^{-} \right| j^\mu_{\rm sc} \left| p \right\rangle
\right)\label{eq:n_sigmaminus_pi0}
\end{eqnarray}
\end{widetext}
for the fourth, sixth and seventh rows of eq.
(\ref{eq:matrix_matrix_elements}), respectively. Notice that the
relationships given in eqs. (\ref{eq:n_sigma0_piminus}) and
(\ref{eq:n_sigmaminus_pi0}) are fully consistent with the relation
given previously in eq. (\ref{eq: amp_sig0_pim_vs_sigm_pi0}).

Finally, it is worth warning the reader that these relations between
the amplitudes are exact in the SU(3) limit, but when one uses the
different physical masses of the involved particles, there will be
SU(3) or SU(2) breaking effects. Nonetheless, these relations can be
used to check that the $\mathcal{A}^{N\rightarrow Y\pi}_i$ constants
of the tables \ref{tab:constants_diagrams} and
\ref{tab:constants_diagrams_resonances} satisfy them. However, one has
to be careful when checking these $\mathcal{A}^{N\rightarrow Y\pi}_i$
constants in some Born diagrams, where there are additional factors
hidden in the standard definitions of the $f^{NY}_i(q^2)$ and
$g^{NY}_1(q^2)$ form factors of tables \ref{tab:vector_ff} and
\ref{tab:axial_ff}.

\end{document}